\documentclass[12pt]{iopart}

\usepackage{graphicx}
\usepackage{amssymb}
\usepackage{amsthm}
\usepackage{color}
\usepackage{epstopdf}
\usepackage{bm}

\begin{document}

\def\cM{\mathcal{M}}
\def\cH{\mathcal{H}}
\def \cZ {\mathcal{Z}}
\def\e{\mathrm{e}}
\def\i{\mathrm{i}}
\def\d{\mathrm{d}}
\renewcommand{\Re}{\mathop{\mathrm{Re}}\nolimits}
\renewcommand{\Im}{\mathop{\mathrm{Im}}\nolimits}
\newcommand{\avg}[1]{\left\langle #1 \right\rangle}
\newcommand{\bra}[1]{\langle#1|}
\newcommand{\ket}[1]{|#1\rangle}
\newcommand{\Ord}{\mathop{\mathcal{O}}\nolimits}
\newcommand{\bras}[2]{{}_{#2}\langle{#1}|}

\definecolor{dgreen}{rgb}{0,0.8,0}
\newcommand{\dgreen}{\color{dgreen}}
\newcommand{\rev}[1]{{\color{red}#1}}
\newcommand{\blue}[1]{{\color{blue}#1}}
\newcommand{\REV}[1]{\textbf{\color{red}#1}}
\newcommand{\BLUE}[1]{\textbf{\color{blue}#1}}
\newcommand{\GREEN}[1]{\textbf{\color{dgreen}#1}}

\title[Statistical distribution of the local purity in a large  quantum system]  
{Statistical distribution of the local purity in a large  quantum system}

\author{A. De Pasquale$^{1,2}$, P. Facchi$^{3,4}$, V. Giovannetti$^{2}$, G. Parisi$^{6,7,8}$,  S. Pascazio$^{1,4}$, A. Scardicchio$^{9,10}$}

\address{$^1$Dipartimento di Fisica and MECENAS, Universit\`a di Bari, I-70126 Bari, Italy}
\address{$^2$NEST, Scuola Normale Superiore and Istituto Nanoscienze-CNR, Piazza dei Cavalieri 7, I-56126 Pisa, Italy }
\address{$^3$Dipartimento di Matematica and MECENAS, Universit\`a di Bari, I-70125 Bari, Italy}
\address{$^4$INFN, Sezione di Bari, I-70126 Bari, Italy}
\address{$^6$Dipartimento di Fisica, Universit\`{a} di Roma ``La Sapienza", 00185 Roma, Italy}
\address{$^7$Centre for Statistical Mechanics and Complexity (SMC), CNR-INFM, 00185 Roma, Italy}
\address{$^8$INFN, Sezione di Roma,  00185 Roma, Italy}
\address{$^9$Abdus Salam International Center for Theoretical Physics, I-34014 Trieste, Italy}
\address{$^{10}$INFN, Sezione di Trieste, I-34014 Trieste, Italy}

\begin{abstract}
The local purity of large many-body 
quantum systems can be studied by following a statistical mechanical approach based on a random matrix model. 
Restricting the analysis to  the case of global pure states,
this method proved to be successful  and a full characterization  of the statistical properties of the local purity 
was obtained by computing  the partition function of the problem. Here 
we generalize these techniques  to the case of global mixed states. In this context, by  uniformly sampling the phase space of states with assigned global mixedness, 
we determine the exact expression of the first two moments of the local purity and a general expression for the moments of higher order. This generalizes previous results, obtained for
globally pure configurations. 
Furthermore, through the introduction of a partition function for a suitable canonical ensemble, we compute the approximate expression of the first moment of the marginal purity in the high temperature regime. 
In the process, we establish 
a formal connection with the theory of quantum twirling maps that provides an alternative,
possibly fruitful, way of performing the calculation.
\end{abstract}


\maketitle

\section{Introduction}

In quantum mechanics the purity of a quantum state measures the amount 
 of indeterminacy 
which originates either from   lack of information on the process that led to its
 preparation, or from
the correlations that have been established with some external party  (environment). Mathematically speaking  
the two mechanisms are strongly interconnected, as the former can always be represented as an instance of the latter 
 via purification~\cite{nielsen, benenticasati}.
For composite quantum system one can distinguish between global purity (i.e. the
purity of the system as a whole), and the local (or ``marginal") purities (i.e the purities  associated with the subsystems that compose it). 
Clarifying the connection between the global and local purities of a quantum state is an important problem of quantum information theory~\cite{MIX2,MIX4,MIX3,MIX1} which is closely related to the characterization of bipartite entanglement~\cite{ENTMEAS,ENTMEAS1}.  In particular, for pure global states, the local purity of a subsystem provides a  direct measure of the bipartite entanglement between the two parts: the smaller 
the purity, the larger the bipartite entanglement. For mixed (non pure) 
global states instead, the connection between local purity and entanglement is more subtle: no
direct relation between the two quantities exists  and bipartite entanglement measures  for the global system {can be} obtained  only by  taking proper \emph{convex roof}  extensions of the local purity
(the average being computed
over {\em all} convex decompositions of the initial global state). Still, studying how the global indeterminacy of a composite system affects the indeterminacy of its
constituents is important on its own and raises fundamental theoretical questions which are deeply interwoven with thermodynamical issues
and  call for a thoughtful investigation.
Specifically here  we focus on the following basic question: 
 given a certain level of global mixedness of a many-body quantum system (say a mixture of gases at thermal equilibrium at  a given temperature), which portion of such indeterminacy  can be ``accounted for" by its constituents? (that is how much the local mixedness contributes to the global one?).
While for classical systems  global and local mixedness are strongly interconnected (e.g., the former is always greater than the latter), in quantum mechanics the relation is more ambiguous, as qualitatively different sort of correlations can be established between the various subsystems.

In order to account for the  large spectrum of possibilities, we resort to a statistical approach, by exploiting tools and techniques imported from classical 
statistical mechanics. The distribution of the local purity of pure global states of large composite systems was studied in Refs.~\cite{matrix,metastable} by using statistical mechanical 
methods. Here we endeavor to generalize the same techniques to a system in a global mixed state. In particular,  by uniformly sampling the states with a given value of the global purity, we determine the exact expression of the first two moments of the local purity and obtain a general formula, valid for arbitrary moments. These findings generalize previous results obtained for
globally pure configurations by  Lubkin~\cite{Lubkin}, Page~\cite{page}, Lloyd and Pagels~\cite{LLoyd}, Scott and Caves~\cite{Scott}, and Giraud~\cite{Giraud,Giraud1}.
Then, through the introduction of a partition function for a suitable canonical ensemble endowed with a Lagrange multiplier, that plays the role of a fictitious temperature, we compute the approximate expression of the first moment of the marginal purity in the high temperature regime. We also establish the scaling of all these quantities  with the dimension of the system, in the thermodynamic limit. Furthermore, an interesting connection appears between our problem and the theory of quantum channels. More precisely, the symmetry properties of the twirling transformations~\cite{tw1} can be proved to be very useful for the computation of the exact expression of the first moment of the local purity. As a final remark,  let us stress the key role played by the introduction of a partition function for the system: it  enables us to translate our problem, and thus the results we found, in terms of the even more general context of the theory of random matrices.

%In particular a partition function for a suitable canonical ensemble was introduced and the problem was translated in terms of the distribution of the eigenvalues of random matrices. Here  we endeavor to generalize the above procedure to a system in a global mixed state. In such a case, however, the direct computation of the partition function is far more complicated, since the average over the unitary group does not factorize. On the contrary, a direct characterization of the first moments associated with the local purity distributions, appears to be feasible.  Therefore, in an attempt to collect some preliminary insight on the statistical properties  of the problem, we focus on this latter issue. 

This paper is organized as follows. In Sec.~\ref{sec:introductionstatisticalapproach} we introduce the notation and set  the basis of the statistical mechanical approach to the problem, starting from the simpler case of pure states and generalizing it to the case of generic mixed states.
In Sec.~\ref{sec:moments} we consider the high temperature expansion of the partition function and
compute the first moments of the purity, by making use of Zuber's
solution of some basic integrals over the unitary group \cite{Zuber}. In Sec.~\ref{sec:tw} we establish the connection  between our problem and the properties of the twirling maps.  We conclude in Sec.~\ref{sec:concl} by summarizing our findings and discussing them in terms of future perspectives.

\section{The statistical approach: partition function} \label{sec:introductionstatisticalapproach}

In this section we review the  statistical approach introduced in Ref.~\cite{matrix} for studying the local
mixedness properties of pure states and discuss its generalization to the case of mixed states.

Consider a bipartite system $X=AB$ made up of two subsystems $A$ and $B$,  described by the Hilbert space $\mathcal{H}_X=\mathcal{H}_A \otimes \mathcal{H}_B$, with $\dim \mathcal{H}_A=N_A$, $\dim \mathcal{H}_B=N_B$ and $\dim \cH_X=N=N_A N_B$. Without loss of generality we will assume that $N_A \leq N_B$. The states of $X$  are represented by the set $\mathfrak{S}({\cal H}_X)$ of nonnegative unit-trace operators (density matrices) on the Hilbert space $\mathcal{H}_X$. The purity of such states, defined as 
\begin{eqnarray}
\pi_{AB}(\rho) := \mathrm{Tr}\rho^2\in [1/N, 1]\;, \label{purityglobal}
\end{eqnarray} 
 for each   $\rho \in \mathfrak{S}({\cal H}_X)$,
 provides  a characterization of the global mixing of the system and induces
a partition of $\mathfrak{S}({\cal H}_X)$ into a collection of distinct subsets $\mathfrak{S}_{x}({\cal H}_X) := \{ \rho \in \mathfrak{S}({\cal H}_X): \mathrm{Tr}\rho^2=x\}$.
The minimum value of  $x =1/N$ is attained  when $X$ is in the completely mixed  state $I/N$,  
whereas the maximum $x=1$ is attained over the set $\mathfrak{S}_{1}({\cal H}_X)$ consisting of all pure states
 $\ket{\psi}_X$.
 % whose
%  corresponding density matrices  $\rho$ are described by rank one  projection operators $\ket{\psi}_X\bra{\psi}$
%   (in this case  one eigenvalue of $\rho$ equals $1$ and all the others vanish). 
   For each   $\rho \in \mathfrak{S}({\cal H}_X)$ we can also define its $A$-{\em local} and $B$-{\em local} purity functions as 
   \begin{eqnarray}
\pi_{A}(\rho):= \Tr \rho_A^2
% =\sum_{j=1}^{N_A} \lambda_{A, j}^2 \in [ 1/N_A,1] 
\;,  \qquad  
% \nonumber \\
 \pi_{B}(\rho) : =\Tr \rho_B^2
  %=\sum_{j=1}^{N_B} \lambda_{B, j}^2  \in [ 1/N_B,1] 
   \;,
\label{eq:purityN_Aloc}
\end{eqnarray}
 with  $\rho_A=\Tr_B \rho$ and $\rho_B=\Tr_A \rho$  being the reduced density matrices of the subsystem $A$ and $B$, respectively.
 
\subsection{Total system in a pure state} \label{sec:introductionstatisticalapproach1} 
 
%To clarify the connection between the global and local purities of a state is an important problem of quantum information theory~\cite{MIX2,MIX3,MIX1}, that is closely  related to the characterization of bipartite entanglement~\cite{ENTMEAS,ENTMEAS1}.  
On the special  set  $\mathfrak{S}_{1}({\cal H}_X)$ of pure states $\rho=\ket{\psi}_X \bra{\psi}$ of $X$, the $A$ and the $B$-local purities coincide 
\begin{equation}
\pi(\psi):= \pi_{A}(\ket{\psi}_X \bra{\psi})=\pi_B(\ket{\psi}_X \bra{\psi}) \;, 
% \Tr \rho_A^2=\Tr \rho_B^2 \;, 
\label{eq:purityN_A}
\end{equation}
and provide a  measure of the bipartite entanglement between $A$ and $B$: the smaller 
$\pi(\psi)$, the larger the entanglement contained in $\ket{\psi}_X \bra{\psi}$.
The statistical distribution of $\pi(\psi)$ on $\mathfrak{S}_{1}({\cal H}_X)$ has been studied in Refs.~\cite{matrix,metastable}.
This was done by introducing the  partition function  
\begin{equation}
\mathcal{Z}(\beta)=
\int
\d\mu (\psi)\; \mathrm{e}^{- \beta\; \pi(\psi)} \;,
\label{eq:pure}
\end{equation}
where the local purity $\pi(\psi)$ of  $|\psi\rangle_X$ plays the role of an effective energy of the system, $\beta$ is a Lagrange multiplier that fixes the value of the purity/energy and selects an isopurity manifold \cite{Kus01}, and
$\d\mu (\psi)$ is a (normalized) measure on the space of pure states $\mathfrak{S}_{1}({\cal H}_X)$.
The natural choice  for the latter %
is induced by the Haar (probability) measure $\mathrm{d}\mu_H(U)$ on the unitary group  $\mathcal{U}(\mathcal{H}_X)\simeq\mathcal{U}(N)$, through the mapping 
\begin{equation}
\label{eq:unitaryrot}
\ket{\psi}_{X}:=U_{X}\ket{\psi_0}_{X}\;,
\end{equation}
with $\ket{\psi_0}_X$ an arbitrary reference unit vector of $\cH_X$. Thus the partition function becomes
\begin{equation}\label{purePartFunction}
\mathcal{Z}(\beta)=\int \d \mu_{H}(U_X)\exp\left(-\beta\Tr(\Tr_B (U_{X}\ket{\psi_0}_{X}\bra{\psi_0}U_{X}^\dagger)^2)\right).
\end{equation}
Then, by noticing that for every $U_X \in \mathcal{U}(N)$ the reduced density matrix $\rho_A=\Tr_B(\ket{\psi}_{X}\bra{\psi})$
can be written as $\rho_A=U_A \Lambda_A U_A^{\dagger}$, with $U_A \in \mathcal{U}(N_A)$ and $\Lambda_A= \mathrm{diag} \{ \lambda_{A, 1}, \lambda_{A, 2}, \ldots \lambda_{A, N_A} \}$, the expression~(\ref{purePartFunction}) 
  becomes
 \begin{equation}\label{eq:pure1}
\fl \qquad \cZ(\beta)
=\int \d \mu_{H}(U_A) \int \d \sigma(\Lambda_A)\;  \e ^{-\beta\Tr \Lambda_A^2}
=\int \d \sigma(\Lambda_A)\;  \e ^{-\beta\Tr \Lambda_A^2} ,
\end{equation}
where 
we exploited  the fact that the measure induced over the density matrices $\rho_A$ by 
$\d \mu_H(U_X)$
factorizes into the product of a measure over the unitary group $\d \mu_{H}(U_A)$ (related to the eigenvectors of $\rho_A$) and  a measure  $\d \sigma(\Lambda_A)$ over the $(N_A -1)$-dimensional simplex of its eigenvalues $\lambda_{A,j}$~\cite{Zyczkowski1, Zyczkowski2}.
In particular, it can be shown that \cite{LLoyd,page,Zyczkowski3}
\begin{eqnarray}\label{eq:mis_autov_A}
\fl\qquad \d \sigma(\Lambda_A)=  C_{N_A,N_B}&&\delta\left(1-\sum_{1\leq i \leq N_A} \lambda_{A,i}\right) \prod_{1\leq j \leq N_A} \theta(\lambda_{A,j})\lambda_{A,j}^{N_B-N_A} 
\nonumber \\
&& \times \prod_{1\leq l<m \leq N_A}(\lambda_{A,l}-\lambda_{A,m})^2 \d ^ {N_A} \lambda_A\;,
\end{eqnarray}

with 
\begin{equation}
C_{N_A,N_B}=\frac{\Gamma(N_A N_B)}{ \prod_{0 \leq j \leq N_A-1}\Gamma(N_B-j) \Gamma(N_A-j+1)},
\end{equation}
and $\theta(x)$ and $\Gamma$ being the unit step and the Euler gamma function, respectively.
 The derivatives  of $\cZ(\beta)$, evaluated for $\beta=0$, yield
 the moments of $\pi(\psi)$ with respect to the measure $\d \mu(\psi)$, i.e.
 \begin{eqnarray} \label{moma}
 {\cal M}_n: = \left\langle\left(\mathrm{Tr} \rho_A^2\right)^n \right\rangle_{\mathfrak{S}_{1}({\cal H}_X)} =
 \int \d\mu(\psi) \; \pi(\psi)^n = 
 (-1)^n \; \left. \frac{\partial^n  \cZ(\beta)}{\partial \beta^n }\right|_{\beta=0} \;.
 \end{eqnarray} 
 These functions fully determine the statistical distribution of $\pi(\psi)$ on $\mathfrak{S}_{1}({\cal H}_X)$ and,
 in the high temperature regime, provide an expansion of $\cZ(\beta)$. 
 More generally, 
 in analogy to what is commonly done in statistical physics,  
 one can also define 
 the moments of $\pi(\psi)$ for $\beta\neq 0$ as  
 \begin{eqnarray}
 {\cal M}_n(\beta): =   \left\langle\left(\mathrm{Tr} \rho_A^2\right)^n \right\rangle_{\beta, \mathfrak{S}_{1}({\cal H}_X)}  = 
 \label{ffd1} 
  \int \d\mu_\beta(\psi) \; \pi(\psi)^n = 
 \frac{(-1)^n}{\cZ(\beta) } \;  \frac{\partial^n  \cZ(\beta)}{\partial \beta^n } \;,
 \end{eqnarray} 
with $\d \mu_\beta(\psi)$ being the canonical measure
\begin{eqnarray}\label{ffhh}
\d \mu_\beta(\psi) :=\d \mu(\psi) \; \frac{  \e ^{-\beta \;  \pi(\psi)} }{\cZ(\beta)} \;.
\end{eqnarray} 
The latter is a deformation of the Haar measure $\d \mu(\psi)$ obtained by including a nonuniform weight 
which explicitly depends upon the local purity, through $\beta$. In particular, as $\beta$ increases 
$\d \mu_\beta(\psi)$ enhances the role of the states with lower values of $\pi(\psi)$ (i.e. larger values of bipartite entanglement), to the extent that
for $\beta \rightarrow +\infty$  only the maximally entangled elements of  $\mathfrak{S}_{1}({\cal H}_X)$ contribute to the values~(\ref{ffd1}). Since $\pi(\psi)$ is bounded above, one can also consider negative temperatures, where the role of less entangled states is enhanced,  and when $\beta\to -\infty$ only the separable  elements of  $\mathfrak{S}_{1}({\cal H}_X)$ contributes to the values~(\ref{ffd1}).
Consequently, across different ranges of temperatures, the moments in Eq.~(\ref{ffd1}) characterize the statistical distribution of the local purity of $X$ computed with respect to a {\em canonical} ensemble whose constituents are selected according  to an effective  thermal distribution
characterized by the parameter $\beta$. 

In the limit of large $N$, the $\beta$-dependence of the statistics of the local purity $\pi(\psi)$ (and hence of the bipartite entanglement of the system) was
characterized in Ref.~\cite{matrix,metastable} by identifying the class of states which maximize the distributions (\ref{ffhh}), i.e.\ typical states with respect to the canonical measure (\ref{ffhh}). In this context it was shown that the system undergoes two main phase transitions, related to different distributions of the eigenvalues $\Lambda_{A}$ of the typical states: a second-order phase transition, mentioned above, associated to a ${\mathbb Z}_2$ symmetry breaking, and related to the vanishing of some eigenvalues of $\rho_A$, followed by a first-order phase transition, associated to the evaporation of the largest eigenvalue from the sea of the others.

\subsection{Total system in a mixed state} \label{sec:introductionstatisticalapproach2} 

A natural question is what happens when the global
system $X$ is in a \emph{mixed} state $\rho$ of purity  $x < 1$, rather than in a pure state. 
A generalization of Eq.~(\ref{eq:pure}) is obtained by replacing $\pi(\psi)$ with  (say)
the $A$-local purity 
$\pi_A(\rho)$ of Eq.~(\ref{eq:purityN_Aloc})  and 
the measure $\d \mu(\psi)$ with a proper measure $\d\mu_{x}(\rho)$ on the set $\mathfrak{S}_{x}({\cal H}_X)$.
This yields the following definition of the partition function of the $A$-local purity 
\begin{equation}%\label{eq:MixedpartitionFunction0}
\mathcal{Z}_A(x, \beta)=\int \mathrm{d} \mu_{x} (\rho) \;
\mathrm{e}^{-\beta\; {\pi}_A(\rho)}
%(\Tr \rho_A^2)}
 = C_{x} \int \mathrm{d}\mu
(\rho) \; \delta(\Tr\rho^2 -x) \;
\mathrm{e}^{-\beta\; {\pi}_A(\rho)}\;,
 \label{eq:mixed}
\end{equation}
where $\mathrm{d}\mu (\rho)$ is a probability measure on the set of mixed states (see below),
and $C_{x} : = \left[ \int \mathrm{d}\mu (\rho) \; \delta(\Tr\rho^2 -x)\right]^{-1}$ a normalization factor. An analogous expression 
for the $B$-local purity partition function $\mathcal{Z}_B(x, \beta)$  is obtained  by replacing $\pi_A(\rho)$ with $\pi_B(\rho)$ in Eq.~(\ref{eq:mixed}): 
notice, however,  that, at variance with the case  analyzed in Ref.~\cite{matrix,metastable}, for  $x < 1$ the partition function $\mathcal{Z}_B(x, \beta)$ will in general differ form   $\mathcal{Z}_A(x, \beta)$. 

It is worth stressing that the  function $\mathcal{Z}_A(x, \beta)$  provides only statistical information on the  local mixedness  of $X$, but not directly on its  bipartite entanglement properties: this is  due to the fact that
for generic mixed states $\rho$ of $X$ the local purities $\pi_A(\rho)$ and $\pi_B(\rho)$ are not entanglement measures. A generalization of Eq.~(\ref{eq:pure}) that retains the ability of characterizing the statistical properties of the
bipartite entanglement of $X$ for $x<1$  could in principle be constructed by
replacing $\pi(\psi)$ with 
 the  convex-roof counterpart of $\pi_A(\rho)$, namely 
$\bar{\pi}_{A}(\rho) = \max_{{\cal E}} \; \sum_j p_j\;  \pi_{A} (\psi_j)$,
where the maximum is taken over all ensembles ${\cal E}:=\{ p_j,  |\psi_j\rangle_{X}\}_j$ which yield a convex decomposition of the mixed state $\rho$ (see e.g.\ Refs.~\cite{BENNETT,CORNELIO}).
The quantity $\bar{\pi}_{A}(\rho)$ is a  proper measure of the bipartite entanglement, 
but  the resulting partition function 
does not allow for a simple analytic treatment and will not  be discussed in the present paper. 

Finally, since there is no unique measure  on mixed states \cite{Zyczkowski3}, we need to properly specify the choice of  $\d \mu(\rho)$ 
which enters in Eq.~(\ref{eq:mixed}) -- the consistency requirement  that  for pure states Eq.~(\ref{eq:mixed})  should reduce back to Eq.\ (\ref{eq:pure})  (i.e.\  $\mathcal{Z}_A(1, \beta)=\mathcal{Z}(\beta)$) does not eliminate such ambiguity. Indeed, as previously discussed for $\rho_A$,  the Hermitian matrix $\rho$ can always be diagonalized by a unitary operation, and as consequence we can write the measure $\mathrm{d}\mu (\rho)$ as the product of a measure on the $(N-1)$-dimensional symplex of the eigenvalues and a measure on the unitary group $\mathcal{U}(N)$ related to the eigenvectors. However, if on the one hand it would be natural to take the Haar measure on $\mathcal{U}(N)$ so that $\mathrm{d}\mu (\rho)=\mathrm{d}\mu (U_X \rho U_X^{\dagger})$, on the other hand the measure on the eigenvalues can be chosen in different possible ways \cite{Slater}. In order to overcome this ambiguity we will use a balanced purification strategy. Let us introduce hence the composite Hilbert space
$\mathcal{H}_{X X'}=\mathcal{H}_X\otimes \mathcal{H}_{X'}$, where
$\mathcal{H}_{X} \simeq \mathcal{H}_{X'}$ are isomorphic. In this $N^2$-dimensional Hilbert space, each $\rho$ of $X$ can be represented by those pure states
 $\ket{\Psi}_{X X'}$  which  provide a purification for such density matrix, i.e. which satisfy the identity 
\begin{equation} \rho=\mathrm{Tr}_{X'}
(\ket{\Psi}_{X X'}\bra{\Psi})\;.
\end{equation}
Thanks to this identification we can now induce a measure on  $\mathfrak{S}({\cal H}_X)$ by sampling 
the pure states on $\mathcal{H}_{X X'}$ according to the unique,
unitarily invariant Haar measure which, as usual,  is induced  by the Haar measure on the unitary group
$\mathcal{U}(N^2)$ through the mapping $\ket{\Psi}_{X X'}:= U_{X X'}\ket{\Psi_0}_{X X'}$, where $\ket{\Psi_0}_{X X'}$ is an arbitrary reference vector and $U_{X X'}\in\mathcal{U}(N^2)$. With this choice the partition function becomes
%the measure 
%$\d \mu_x(\rho)$ of 
%Eq.~(\ref{eq:mixed}) gets replaced by 
%\begin{eqnarray}
%\d \mu_x(\rho) \rightarrow \mathrm{d}\mu_{H}(U_{X X'})\,
%\delta\left(x^2-\Tr(\Tr_{X'}(\ket{\Psi}_{X X'}\bra{\Psi})^2\right)
%\end{eqnarray} 
%
%our partition function~(\ref{eq:mixed}) becomes
\begin{eqnarray}
\fl \mathcal{Z}_A(x, \beta)= C_x \int
\mathrm{d}\mu_{H}(U_{X X'})\,
\delta\left(x-\Tr(\Tr_{X'}(\ket{\Psi}_{X X'}\bra{\Psi})^2\right)
\mathrm{e}^{-\beta\; \Tr
\left(( \Tr_B( \Tr_{X'} \ket{\Psi}_{X X'}\bra{\Psi}))^2\right) },
\label{eq:ZUN2}
\nonumber\\
\end{eqnarray}
where we used the fact that 
$\rho_A=\mathrm{Tr}_B\rho=\mathrm{Tr}_B(\mathrm{Tr}_{X'}
\ket{\Psi}_{X X'}\bra{\Psi})$. Analogously to what we have seen for the pure case, $x=1$, by writing
$\rho=U_X \Lambda_X U_X^\dagger$ with 
%\begin{equation}
$\Lambda_X=\mathrm{diag}(\lambda_1, \lambda_2, ..., \lambda_N)$,
%\end{equation}
we get
\begin{eqnarray}\label{eq:partition function unitary rotated}
\fl\qquad \mathcal{Z}_A(x, \beta)= C_x \int
\mathrm{d}\mu_H(U_X) \int \mathrm{d}\sigma(\Lambda_{X})\; \delta\Big(x-\Tr \Lambda_X^2\Big)\; \mathrm{e}^{-\beta\;
\mathrm{Tr}
\left((\mathrm{Tr}_B(U_X\Lambda_X {U_X}^\dagger))^2\right)}, 
\end{eqnarray}
where  $\mathrm{d}\mu_H(U_X)$ is the Haar measure on $\mathcal{U}(N)$ 
and
\begin{equation}\label{eq:mixed1}
\mathrm{d}\sigma(\Lambda_{X})=C_N\;  \delta\left(1-\sum_{1\leq i \leq N}\lambda_i\right)\; \prod_{1 \leq i \leq N} \theta(\lambda_i)
\prod_{1\leq i<j \leq N}(\lambda_i-\lambda_j)^2 \mathrm{d}^N \lambda,
\end{equation}
with
\begin{equation}
C_{N}=\frac{\Gamma(N^2)}{\Gamma(N+1) \prod_{1\leq k \leq N } \Gamma(k)^2}.
\end{equation}
Therefore, we have identified the measure $\d \mu_x(\rho)$ of Eq.~(\ref{eq:mixed}) with  
\begin{eqnarray}\label{defdmu}
\d \mu_x(\rho) = C_x \mathrm{d}\mu_H(U_X) \mathrm{d}\sigma(\Lambda_{X})\delta\Big(x-\Tr \Lambda_X^2\Big)\;.
\end{eqnarray} 
Notice that in the case of pure states, i.e.\ $x=1$, the density operator of the system reduces to $\rho=\ket{\psi}_{X}\bra{\psi}$, where $\ket{\psi}_{X}=U_{X}\ket{\psi_0}_{X}$, $\ket{\psi_0}_X$ being an arbitrary reference state (see Eq. (\ref{eq:unitaryrot})), and  the matrix $\Lambda_{X}$ becomes a rank one projection. Thus, the expression (\ref {eq:partition function unitary rotated}) reduces to  (\ref{purePartFunction}), namely
\begin{eqnarray}
\cZ_A(1, \beta)=\cZ(\beta)\;. \label{identita}
\end{eqnarray} 

%%%%%%%% Tagliato %%%%%%%%%%%%%
%
%To make the dependence upon the variable of integration more explicit it is worth introducing 
%the spectral decomposition of $\Lambda_X$, i.e. 
%\begin{equation}
%\Lambda_X=\sum_{1\leq i \leq N} \lambda_i P_i, \quad \lambda_i \in [0,1]
%\end{equation}
%where  $P_i=|i\rangle_X\langle i|$ with $i=1, \ldots, N$  are the projections belonging to the reference basis
%$\{\ket{i}_X\}\subset \mathcal{H}_X$, that is $P_i=(\delta_{k,i} \delta_{i,l})_{1\leq k, l \leq N}$.  Defining the reduced density matrix $\rho_{Ai}=\rho_{Ai}(U_X)=\mathrm{Tr}_B(U_XP_i {U_X}^\dagger)$, associated to each eigenstate $U_X\ket{i}_X$ of $\rho$, the partition function becomes then
%\begin{equation}
% \mathcal{Z}_A(x,\beta)
% %&=&\int \mathrm{d}\mu_H(U_X) \mathrm{d}\sigma(\Lambda_{X})\delta\Big(x^2-\sum_i \lambda_i^2\Big)\mathrm{e}^{-\beta(\mathrm{Tr}_A (\sum_{i,j}\lambda_i\lambda_j \rho_i
% %\rho_j)}\nonumber \\
% = C_x\int \mathrm{d}\mu_H(U_X) \int \mathrm{d}\sigma(\Lambda_{X})
% \delta\Big(x-\sum_i \lambda_i^2\Big)
% \mathrm{e}^{-\beta\sum_{i,j}\lambda_i\lambda_j
% \pi_{ij}(U_X)}\;,
% %\nonumber\\
%% &=&\left\langle\mathrm{e}^{-\beta
% %\sum_{i,j}\lambda_i\lambda_j\pi_{ij}}\right\rangle,
% %\simeq 1 - \beta\left\langle\sum_{i,j}\lambda_i\lambda_j\pi_{ij}\right\rangle \Bigg|_{\beta=0}+ ..
%\label{eq:main}
% \end{equation}
% where $\pi_{ij}(U_X)= \mathrm{Tr}_A(\rho_{Ai}\rho_{Aj})$.
% %and where we introduce the notation 
% %$\langle \cdots \rangle$ to 
% %indicate the average over $\lambda_j$ induced by $ \mathrm{d}\mu_H(U_X) \; \mathrm{d}\sigma(\Lambda_{X})$.
%
%%%%%%%%%%%%%%%%%%%%%

\subsection{Asymptotic behavior and analysis of moments} \label{sec:Asymptotic behavior and moment analysis}

For $x < 1$, the integration over the unitary group $\mathcal{U}(N)$ in Eq.~(\ref{eq:partition function unitary rotated})
%(\ref{eq:main})
 does not factorize, making the computation of the partition function far more complicated than for the case of a pure state~(\ref{eq:pure1}). 
 The only notable exception is the case of maximally mixed states (i.e. $x = 1/N$), when the Dirac delta in 
 Eq.~(\ref{eq:partition function unitary rotated}) selects a unique diagonal matrix $\Lambda_X$ (the totally mixed state of $X$).
 This makes the exponent equal to  $\mathrm{e}^{-\beta/N_A}$ for all $U_X$ and yields the following exact expression 
\begin{eqnarray}\label{ffhh11}
\cZ_A(1/N,\beta) = \mathrm{e}^{-\beta/N_A}\;.
\end{eqnarray}
Otherwise, for intermediate values of the purity, $1/N < x < 1$, the situation is much more complicated.  Still, as we will show in the following, at small $\beta$ the evaluation of  the moments ${\cal M}^{A}_n(x, \beta)$ associated with $\cZ_A(x,\beta)$  admits an exact analytical treatment. The latter are
formally defined as
\begin{equation}
 {\cal M}^{A}_n(x, \beta): =  \int \d \mu_{x,\beta}(\rho) \; \pi_A^n(\rho) = \label{defma}
 % \left\langle\left(\mathrm{Tr}_A \rho_A^2\right)^n \right\rangle_{\beta, \mathfrak{S}_{x}({\cal H}_X)}  = 
% \label{ffd1} 
%  \int \d\mu_\beta(\rho) \; \pi_A^n(\rho) = 
 \frac{(-1)^n}{\cZ_A(x,\beta) } \;  \frac{\partial^n  \cZ_A(x,\beta)}{\partial \beta^n } \;,
 \end{equation} 
and represent the average value of $\pi_A^n(\rho)$ with the canonical measure
\begin{eqnarray}\label{eq:mubeta}
\d \mu_{x,\beta}(\rho) := \d \mu_x(\rho)\; \frac{  \e ^{-\beta \;  \pi_A(\rho)} }{\cZ_A(x,\beta)} \;,
\end{eqnarray} 
with $\d \mu_x(\rho)$ given by Eq.~(\ref{defdmu}). 
For pure states ($x=1$) the ${\cal M}^{A}_n(x, \beta)$ coincide with the moments ${\cal M}_n(\beta)$ defined in Eq.~(\ref{ffd1}):  at $\beta=0$,
in the large $N$ limit the expression for such quantities has been computed in Ref.~\cite{matrix}, while the exact expressions for  first five  of them 
 can be found  in  Ref.~\cite{Giraud}.
 In the case of a totally mixed state  ($x=1/N$) Eq.~(\ref{ffhh11}) yields instead values which are independent of the
temperature $\beta$, namely 
\begin{eqnarray} \label{kkll}
 {\cal M}^{A}_n(1/N, \beta) =  {\cal M}^{A}_n(1/N, 0) = {N_A}^{-n} \;.
\end{eqnarray}
For intermediate values of $x$, by 
expanding Eq.~(\ref{defma}) up to the first order in $\beta$,  we get  
\begin{equation}
\fl \quad {\cal M}_n^A(x,\beta)  \sim
%&=& 
% {\cal M}_n^A(0,x^2) + \beta \; \left. \frac{\partial}{\partial\beta' }   [ -  \frac{\partial}{\partial \beta'}  \ln \cZ_A(\beta', x^2) 
%]\right|_{\beta'=0} + {\cal O}(\beta^2)\;. \label{exp33}
%\nonumber \\
 {\cal M}_n^A(x,0) - \beta \; [  {\cal M}^A_{n+1}(x,0) - {\cal M}^A_1(x,0)\; {\cal M}^A_{n}(x,0)] %+ {\cal O}(\beta^2) 
\;, \qquad \beta\to 0 .
\label{exp333}
\end{equation} 
Incidentally, 
notice that in agreement with Eq.~(\ref{kkll}),  the $\beta$-corrections of Eq.~(\ref{exp333}) vanish when $x=1/N$. 
The above expression shows that, at least in the high  temperature regime, we can focus
on the unbiased moments  $ {\cal M}^{A}_n(x, 0)$.

\section{Moments of the purity at $\beta=0$}
\label{sec:moments}
\subsection{First moment}\label{sec:Zuber1}
In this section we compute the exact first moment of the purity ${\cal M}_1^A(x,0)$, by making use of Zuber's
solution of some basic integrals over the unitary group \cite{Zuber}. In particular, we will show that the only dependence on the spectrum of the density matrix of the global system is in terms of its purity $x$, whose value is fixed 
in the partition function~(\ref{eq:mixed}). 

Let us fix the spectrum of the global density matrix $\rho$ of the system:
\begin{equation}\label{eq:fixedSpectrum}
\Lambda_X = \mathrm{diag} (\lambda_{N_B(\alpha-1)+ \beta}),
\end{equation}
with $\alpha =1,\dots, N_A$ and $\beta=1, \dots, N_B$. A purification of  $\Lambda_X$ in the space $\cH_{X X'}$, with $X=AB$ and $X'=A'B'$ is
\begin{eqnarray}
|\Psi\rangle_{X X'}  = \sum_{1 \leq \alpha \leq N_A} \sum_{1 \leq \beta \leq N_B} \sqrt{\lambda_{\alpha \beta}}\; \; \ket{\alpha \beta}_{AB}\otimes  \ket{\alpha \beta}_{A'B'} ,
\label{eq.ref_state}\end{eqnarray}
where we have set, for simplicity, $\lambda_{\alpha \beta}=\lambda_{N_B(\alpha-1)+ \beta}$ and $\ket{\alpha \beta}_{A B}=\ket{\alpha}_A\otimes \ket{\beta}_B$,  $\{ \ket{\alpha}_{A}
\}$ and  $\{ \ket{\beta}_{B}\}$  
($\{ \ket{\alpha}_{A'}\}$ and $\{ \ket{\beta}_{B'}\}$) being the reference basis in $\cH_A$ and $\cH_B$ ($\cH_{A'}$ and $\cH_{B'}$), respectively.
The set of vectors in $\cH_{X X'}$ with the same Schmidt coefficients is given by $U_{XX'}\ket{\Psi}_{XX'}$, where $U_{X X'}={U}_{X} \otimes {U}_{X'}$, with $U_X, U_{X'} \in \mathcal{U}(N)$, and yields the set of density matrices with the same spectrum $\Lambda_X$, namely $\rho= U_X \Lambda_X U_X^\dag$. By partial tracing over subsystem $B$ one obtains
the set of reduced density matrices $\rho_A(U)=
 \Tr_{B}(\Tr_{X'}(U_{XX'} \ket{\Psi}_{XX'}\bra{ \Psi}U^\dag_{XX'} ))$. Notice that this expression does not depend on $U_{X'}\in \mathcal{U}(N)$:
 \begin{eqnarray}
 \rho_{A}&=& \sum_{1 \leq \alpha \leq N_A} \sum_{1 \leq \beta \leq N_B} \lambda_{\alpha \beta} \Tr_{B}(U_{AB} \ket{\alpha \beta}_{AB}\bra{\alpha \beta}U_{AB}^{\dagger})\nonumber\\
 &=& \sum_{1 \leq \alpha \leq N_A} \sum_{1 \leq \beta, j \leq N_B}  \lambda_{\alpha \beta} \, \bras{j}{B} U_{AB} \ket{\alpha \beta}_{AB}\bra{\alpha \beta}U_{AB}^{\dagger} \ket{j}_B \;,
 \end{eqnarray}
The purity is given by
  \begin{eqnarray}\label{eq:locpuritydoublespace}
\pi_A(U_X \Lambda_X U^\dagger_X)=
\Tr\rho_A^2=\sum_{1\leq  \alpha_1,\alpha_2\leq N_A}\sum_{1 \leq \beta_1,\beta_2 \leq N_B} \sum_{1 \leq  j_1, j_2 \leq  N_B}   \lambda_{\alpha_1 \beta_1} \lambda_{\alpha_2 \beta_2} \nonumber \\\qquad\qquad\qquad\qquad \times \bras{\alpha_2 \beta_2}{AB}U_{AB}^{\dagger} \ket{j_2}_B\bra{j_1} U_{AB} \ket{\alpha_1 \beta_1}_{AB}\nonumber \\ \qquad\qquad\qquad\qquad\times\, \bras{\alpha_1 \beta_1}{AB}U_{AB}^{\dagger} \ket{j_1}_B\bra{j_2} U_{AB} \ket{\alpha_2 \beta_2}_{AB}\;,
 \end{eqnarray}
which, by the completeness relation for subsystem $A$, becomes
  \begin{eqnarray}\label{Vtre1}
\pi_A(U_X \Lambda_X U^\dagger_X)=\sum_{1\leq  \alpha_1,\alpha_2\leq N_A}\sum_{1 \leq \beta_1,\beta_2 \leq N_B} \sum_{1 \leq  i_1, i_2 \leq  N_A}\sum_{1 \leq  j_1, j_2 \leq N_B}   \lambda_{\alpha_1 \beta_1} \lambda_{\alpha_2 \beta_2} \nonumber \\\qquad\qquad\qquad\qquad \times \bras{\alpha_2 \beta_2}{AB}U_{AB}^{\dagger} \ket{i_1 j_2}_{AB}\bra{i_1 j_1} U_{AB} \ket{\alpha_1 \beta_1}_{AB}\nonumber \\ \qquad\qquad\qquad\qquad\times\, \bras{\alpha_1 \beta_1}{AB}U_{AB}^{\dagger} \ket{i_2 j_1}_{AB}\bra{i_2 j_2} U_{AB} \ket{\alpha_2 \beta_2}_{AB}.
 \end{eqnarray}
 
Let us now compute the first moment of the purity (\ref{defma}) at $\beta=0$.  By recalling that $\cZ_A(x, 0)=1$, $\d \mu_{x,0}(\rho)= \d \mu_x(\rho)$, 
and $\d \mu_x(\rho)= C_x \mathrm{d}\mu_H(U_X) \mathrm{d}\sigma(\Lambda_{X})\delta\Big(x-\Tr \Lambda_X^2\Big)$,
%See Eqs. (\ref{eq:mubeta}) and  (\ref{defdmu}).
we get
\begin{eqnarray}
\fl\qquad \qquad  {\cal M}^{A}_n(x, 0) &=&  \int \d \mu_{x}(\rho) \; \pi_A^n(\rho) 
\nonumber\\
&=& C_x \int \d\sigma(\Lambda_X) \delta(x-\Tr \Lambda^2_X)\;
 {\cal M}^{A}_n(\Lambda_X)
=: \langle {\cal M}^{A}_n(\Lambda_X)\rangle_x,
\label{eq:MAnx0}
 \end{eqnarray} 
where
\begin{equation}
\fl \qquad \qquad {\cal M}^{A}_n(\Lambda_X):= \int \d\mu_H(U_X)\; \pi_A^n(U_X \Lambda_X U_X^\dagger).
\label{eq:MAnLX}
\end{equation}
From~(\ref{Vtre1}) the average over the unitary group of the first moment particularizes to
\begin{eqnarray}
\fl {\cal M}_1^A(\Lambda_X)
%&=&\int \d \mu_H(U_X) \Tr_A \rho_A^2(U) \nonumber\\ 
 &=&\sum_{1\leq  \alpha_1,\alpha_2\leq N_A}\sum_{1 \leq \beta_1,\beta_2 \leq N_B}   \lambda_{\alpha_1 \beta_1} \lambda_{\alpha_2 \beta_2}\nonumber\\
\fl  &&\times \sum_{1 \leq  i_1, i_2 \leq  N_A}\sum_{1 \leq  j_1, j_2 \leq N_B} \int \d \mu_H(U_X) U_{i_1 j_1, \alpha_1 \beta_1}U_{i_2 j_2, \alpha_2 \beta_2} 
%\nonumber\\
%\fl  &&\qquad\qquad\qquad\qquad\qquad\times 
U^\dag_{\alpha_2 \beta_2,i_1 j_2}U^\dag_{\alpha_1 \beta_1,i_2 j_1},
%\nonumber\\ \fl
\label{eq:unint}\end{eqnarray}
where $U_{i j, \alpha \beta}=\bras{i j }{AB}U_{AB} \ket{\alpha \beta}_{AB}$.
This integral can be explicitly computed by using  Zuber's solution \cite{Zuber}:
\begin{equation}
\fl\quad \int \d \mu_H (U) U_{i_1 j_1} \ldots U_{i_n j_n} (U) U_{k_1 l_1}^{\dag} \ldots U_{k_n l_n}^{\dag}=\sum_{\tau, \sigma \in S_n} C[\sigma]\prod_{1\leq a \leq n}\delta(i_a, \ell_{\tau(a)})\delta(j_a,k_{\tau \sigma(a)}), 
\label{eq:tecnicaZuber}
\end{equation}
with
\begin{equation}
C[\sigma]=\sum_{|Y|=n} \frac{(\chi^{(k)}(1))^2\chi^{(k)}([\sigma])}{n!^2 s_k(I)},
\label{eq:tecnicaZuber1}
\end{equation}
where $C[\sigma]$ is the sum over the Young diagrams $Y$ of the character $\chi^{(k)}([\sigma])$  of the symmetric group $S_n$ associated to $Y$, depending on the conjugacy class $[\sigma]$ of the permutation $\sigma$, $s_k(I)$ is the dimension of the representation, 
%of $Y$ in terms of the linear group $\mbox{GL}(N)$, 
and $\delta(a,b)$ is the Kronecker delta. Applying this solution to (\ref{eq:unint}) we get
\begin{eqnarray}
\fl \quad {\cal M}_1^A(\Lambda_X) &=&\sum_{1\leq  \alpha_1,\alpha_2\leq N_A}\sum_{1 \leq \beta_1,\beta_2 \leq N_B}   \lambda_{\alpha_1 \beta_1} \lambda_{\alpha_2 \beta_2}
\nonumber\\ 
\fl &&\times\sum_{\tau, \sigma \in S_2} C[\sigma] f_1(\tau)\delta(\alpha_1 \beta_1,\alpha_{\tau\sigma(2)}\beta_{\tau\sigma(2)})\delta(\alpha_2 \beta_2,\alpha_{\tau\sigma(1)}\beta_{\tau\sigma(1)})\nonumber\\ 
\fl  &=&\sum_{1\leq  \alpha_1,\alpha_2\leq N_A}\sum_{1 \leq \beta_1,\beta_2 \leq N_B}   \sum_{\tau, \sigma \in S_2} \sum_{c\in \mathcal{C}(S_2)} C[\sigma] f_1(\tau) \delta([\tau\sigma s],c) \lambda_{\alpha_{c(1)}\beta_{c(1)}} \lambda_{\alpha_{c(2)}\beta_{c(2)}},
\nonumber\\ 
\fl \label{eq:prim_momZuber}
\end{eqnarray}
where $ f_1(\pi)$ depends on the permutation $\pi \in S_2$
\begin{equation}
\fl \qquad\qquad f_1(\pi):=\sum_{1 \leq i_1,i_2 \leq N_A}\delta(i_1 ,i_{\pi(1)} ) \delta(i_2 ,i_{\pi(2)})\sum_{ 1\leq j_1,j_2 \leq N_B}\delta(j_1, j_{\pi(2)}) \delta(j_2, j_{\pi(1)}),
\end{equation}
$s\in S_2$  is the transposition (swapping) of pairs of nearby indices ($[s]=[2]$)
\begin{eqnarray}\label{eq:s}
i_{s(1)} = i_{2}  \quad  \mbox{and} \quad i_{s(2)} = i_{1}\;,
\end{eqnarray}
and $\mathcal{C}(S_2)=\{[1^2],[2]\}$  is the set of the conjugacy classes of the symmetric group $S_2$. From (\ref{eq:prim_momZuber}) it can be easily inferred that the only possible contributions of the spectrum are related to the conjugacy  classes of the symmetric group $S_2$:
\begin{eqnarray}\label{contrSpectrum1}
\left[\tau\sigma s \right]=\left[1^2\right]& \Rightarrow&\left(\sum_{1\leq \alpha \leq N_A}\sum_{1\leq \beta \leq N_B}\lambda_{\alpha\beta}\right)^2=1,\nonumber\\
\left[\tau\sigma s\right]=\left[2\right]& \Rightarrow& \left(\sum_{1\leq \alpha \leq N_A}\sum_{1\leq \beta \leq N_B}\lambda_{\alpha \beta}^2\right)=\Tr\Lambda_X^2.
\end{eqnarray}
By summing and by using the explicit expressions of the coefficients~(\ref{eq:tecnicaZuber1}) \cite{Zuber}
\begin{equation}
\label{ zubercoeff}
C[1^2]=\frac{1}{(N-1)(N+1)}, \qquad C[2]=-\frac{1}{(N-1)N(N+1)},
\end{equation}
we get
\begin{equation}\label{eq:primMomU}
{\cal M}_1^A(\Lambda_X)=\frac{N_A(N_B^2-1)}{N_A^2N_B^2-1} + \frac{N_B(N_A^2-1)}{N_A^2N_B^2-1} \Tr \Lambda_X^2.
\end{equation}
The first moment of the purity of subsystem $A$, is the average of (\ref{eq:primMomU}) over the spectrum of the system. By plugging~(\ref{eq:primMomU}) into~(\ref{eq:MAnx0}), we finally get 
\begin{equation}\label{eq:firstMom}
{\cal M}_1^A(x,0)
%= \int \d\mu_{x}(\Lambda_X) \; {\cal M}_1^A(\Lambda_X)
= \frac{N_A(N_B^2-1)}{N_A^2N_B^2-1} + \frac{N_B(N_A^2-1)}{N_A^2N_B^2-1} x.
\end{equation} 
%with $\d\mu_{x}(\Lambda_X)$ being any probability measure on the spectrum space which select the input purity equal to $x^2$.
Notice that for $x=1$ this expression reduces to the one given in~\cite{Lubkin,LLoyd,page,Scott,Giraud,Giraud1}.

\paragraph{Some special cases:--} It is worth noticing that for a balanced bipartition $N_A=N_B=\sqrt{N}\gg1$ Eq.~(\ref {eq:firstMom}) yields 
\begin{equation}
{\cal M}_1^A(x,0) = \frac{\sqrt{N} (1+x)}{N+1}\sim \frac{1+x}{\sqrt{N}}\;.
\label{eq:VVFIN}
 \end{equation}
 At $x=1$ (i.e.\ pure global states), Eq.~(\ref{eq:VVFIN}) coincides with that obtained in Refs.~\cite{matrix,Lubkin,Giraud}. 
 Finally, consider the case in which $\rho$ is
 maximally mixed, i.e. is the density matrix $I/N$. In this case $x=1/{N}$ and Eq.~(\ref {eq:firstMom})
 gives
 \begin{eqnarray}
{\cal M}_1^A(1/N,0) =\frac{1}{N_A}\;,
 \end{eqnarray}
 in agreement with the general result~(\ref{kkll}).

\subsection{k-th moment} \label{sec:kthMom}
The technique shown in the previous section can be easily generalized in order to compute from~(\ref{eq:MAnLX})
higher moments  at $\beta=0$. We get
\begin{eqnarray}
\fl{\cal M}_k^A(\Lambda_X)&=&\int \d \mu_H(U_X) \; \pi_A^k(U_X\Lambda_X U_X^\dagger)
\nonumber\\ 
% &=&\sum_{1\leq  \alpha_1,\ldots, \alpha_{2k}\leq N_A}\sum_{1 \leq \beta_1,\ldots,\beta_{2k} \leq N_B}   \lambda_{\alpha_1 \beta_1} \ldots \lambda_{\alpha_{2k} \beta_{2k}}\nonumber\\
% &&\times \sum_{1 \leq  i_1, \ldots, i_{2k} \leq  N_A}\sum_{1 \leq  j_1, \ldots j_{2k} \leq N_B}  \nonumber\\
% && \times\int \d \mu_H(U_X) U_{i_1 j_1, \alpha_1 \beta_1} U_{i_2 j_2, \alpha_2 \beta_2} \ldots U_{i_{2k-1} j_{2k-1}, \alpha_{2k-1} \beta_{2k-1}} U_{i_{2k} j_{2k}, \alpha_{2k} \beta_{2k}} \nonumber\\
% &&\qquad\times U^\dag_{\alpha_2 \beta_2,i_1 j_2}U^\dag_{\alpha_1 \beta_1,i_2 j_1}\ldots U^\dag_{\alpha_{2k} \beta_{2k},i_{2k-1} j_{2k}}U^\dag_{\alpha_{2k-1} \beta_{2k-1},i_{2k} j_{2k-1}}.
% \label{eq:unintk}
%\end{eqnarray}
%\begin{eqnarray}
%\fl{\cal M}_k^A(\Lambda_X)&=&\int \d \mu_H(U_X) \Tr_A \rho_A^2(U)
%\nonumber\\ 
 &=&\sum_{1\leq  \alpha_1,\ldots, \alpha_{2k}\leq N_A}\sum_{1 \leq \beta_1,\ldots,\beta_{2k} \leq N_B}  \prod_{1\leq i \leq 2k} \lambda_{\alpha_i \beta_i} \nonumber\\
 &&\times \sum_{1 \leq  i_1, \ldots, i_{2k} \leq  N_A}\sum_{1 \leq  j_1, \ldots, j_{2k} \leq N_B} \int \d \mu_H(U_X) \prod_{1\leq \ell \leq 2k}U_{i_\ell j_\ell, \alpha_\ell \beta_\ell} \nonumber\\
 &&%\qquad \qquad \qquad \qquad \qquad
  \times \prod_{1\leq m\leq k}U^\dag_{\alpha_{2m} \beta_{2m},i_{2m-1} j_{2m}}U^\dag_{\alpha_{2m-1} \beta_{2m-1},i_{2m} j_{2m-1}} \;.\label{eq:unintk}
 \end{eqnarray}
  Equation~(\ref{eq:tecnicaZuber}) for $n=2 k$ gives
\begin{eqnarray}
\fl{\cal M}_k^A(\Lambda_X) &=& \sum_{1\leq  \alpha_1, \ldots,\alpha_{2k}\leq N_A}\sum_{1 \leq \beta_1,\ldots, \beta_{2k} \leq N_B}   \sum_{\tau, \sigma \in S_{2k}} \sum_{c\in \mathcal{C}(S_{2k})} C[\sigma] f_k(\tau) \delta([\tau\sigma s],c)
\prod_{1\leq i \leq 2k}\lambda_{\alpha_{c(i)}\beta_{c(i)}} , \nonumber \\
\fl \label{eq:kMomUnitIntl}
\end{eqnarray}
where  $ f_k(\pi)$ depends on the permutation $\pi \in S_{2k}$
%\begin{eqnarray}
%f_k(\pi):=&\sum_{1 \leq i_1, \ldots i_{2k} \leq N_A}\delta(i_1,i_{\pi(1)} ) \delta(i_2,i_{\pi(2)})\ldots \delta(i_{2k-1} ,i_{\pi%(2k-1)} ) \delta(i_{2k},i_{\pi(2k)})
%\nonumber\\
%&\sum_{ 1\leq j_1,\ldots j_{2k} \leq N_B}\delta(j_1, j_{\pi(2)}) \delta(j_2,j_{\pi(1)})\ldots \delta(j_{2k-1}, j_{\pi(2k)}) \delta(j_%{2k}, j_{\pi(2k-1)})\nonumber\\
%\end{eqnarray}\label{eq:fk}
\begin{eqnarray}
\fl   f_k(\pi) := \sum_{1 \leq i_1, \ldots, i_{2k} \leq N_A}\sum_{1 \leq j_1, \ldots, j_{2k} \leq N_B}\prod_{1\leq \ell \leq 2k}\delta(i_{\ell} ,i_{\pi(\ell)} )\prod_{1\leq m \leq k}\delta(j_{2m-1} ,j_{\pi(2m)}) \delta(j_{2m} ,j_{\pi(2m-1)}) \nonumber\\
\label{eq:fk}
\end{eqnarray}
and, analogously to Eq. (\ref{eq:s}), $s$ is the swapping of pairs of nearby indices 
%, $i=1, \ldots, N_{A}$ or $N_{B}$:
\begin{equation}\label{eq:sk}
 i_{s(2\ell-1)}=i_{2\ell} \qquad \mbox{and}  \qquad   i_{s(2\ell)}=i_{2\ell-1}  \quad \forall  \, \ell=1,\ldots k.
\end{equation}
Observe that when $k=1$ we retrieve 
${\cal M}_1^A(\Lambda_X)$ (see Eq. (\ref {eq:prim_momZuber})).
The different contributions of the spectrum can be classified in terms of the the conjugacy classes of the symmetric group, as shown in Eq. (\ref{contrSpectrum1}). However, for $k>1$, they do not depend only upon the purity $x=\Tr\Lambda_X^2$, but exhibit a more complex dependence on the spectrum, through
 its higher order invariants $\Tr \Lambda_X^k$, with $k>2$. Thus the integral on the spectrum~(\ref{eq:MAnx0})
%\begin{equation}
%{\cal M}_k^A(x,0)= \int \d\mu_{x}(\Lambda_X) \; {\cal M}_k^A(\Lambda_X)
%\end{equation} 
is in general non trivial. 
\subsection{Second moment}
Now let us fully compute the second moment of the purity for arbitrary bipartite states, with purity $x \in [1/N,1]$,  generalizing some results found for pure states, $x=1$, \cite{matrix,Lubkin,Giraud}.
The second moment can be directly computed by setting $k=2$ in Eq.~(\ref{eq:unintk}). The expression for the coefficients $C[\pi]$ in (\ref{eq:tecnicaZuber}), when $\pi \in \mathcal{C}(S_{4})$  is \cite{Zuber}
\begin{eqnarray}
C\left[1^4\right]&=&\frac{N^4-8N^2+6}{(N-3)(N-2)(N-1)N^2(N+1)(N+2)(N+3)}\nonumber\\
C\left[2,1^2\right]&=&-\frac{1}{(N-3)(N-1)N(N+1)(N+3)}\nonumber\\
C\left[2^2\right]&=&\frac{N^2+6}{(N-3)(N-2)(N-1)N^2(N+1)(N+2)(N+3)}\nonumber\\
C\left[3,1\right]&=&\frac{2N^2-3}{(N-3)(N-2)(N-1)N^2(N+1)(N+2)(N+3)}\nonumber\\
C\left[4\right]&=&-\frac{5}{(N-3)(N-2)(N-1)N(N+1)(N+2)(N+3)}.
\end{eqnarray}
The symmetric group $S_4$, consists of five conjugacy classes, that yield the following contributions to the integral (\ref{eq:kMomUnitIntl}) in terms of the spectrum of $\rho$:
\begin{eqnarray}
 \left[\tau\sigma s\right]=\left[1^4\right]& \Rightarrow&\left(\sum_{1\leq \alpha \leq N_A}\sum_{1\leq \beta \leq N_B}\lambda_{\alpha\beta}\right)^4=1\nonumber\\  
\left[\tau\sigma s\right]=\left[2,1^2\right]& \Rightarrow& \left(\sum_{1\leq \alpha_1 \leq N_A}\sum_{1\leq \beta_1 \leq N_B}\lambda_{\alpha_1\beta_1}^2\right)\left(\sum_{1\leq \alpha_2 \leq N_A}\sum_{1\leq \beta_2 \leq N_B}\lambda_{\alpha_2\beta_2}\right)^2=\Tr \Lambda_X^2\nonumber\\
\left[\tau\sigma s\right]=\left[2^2\right]& \Rightarrow& \left(\sum_{1\leq \alpha \leq N_A}\sum_{1\leq \beta \leq N_B}\lambda_{\alpha\beta}^2\right)^2=(\Tr \Lambda_X^2)^2\nonumber\\
\left[\tau\sigma s\right]=\left[3,1\right]& \Rightarrow& \left(\sum_{1\leq \alpha_1 \leq N_A}\sum_{1\leq \beta_1 \leq N_B}\lambda_{\alpha_1\beta_1}^3\right)\left(\sum_{1\leq \alpha_2 \leq N_A}\sum_{1\leq \beta_2 \leq N_B}\lambda_{\alpha_2\beta_2}\right)= \Tr \Lambda_X^3
%\left(\sum_{\mu_1,\nu_1}\lambda_{\mu_1,\nu_1}^3\right)
\nonumber\\
\left[\tau\sigma s\right]=\left[4\right]& \Rightarrow& 
\sum_{1\leq \alpha \leq N_A}\sum_{1\leq \beta \leq N_B}\lambda_{\alpha\beta}^4=
\Tr \Lambda_X^4 ,
\end{eqnarray}
with $\tau, \sigma \in S_{4}$ and $s \in S_{2}$ defined in (\ref{eq:sk}) being $k=2$. By gathering all
we get 
\begin{eqnarray}
{\cal M}_2^{A}(\Lambda_X)=&&\Bigg\{ \qquad\quad \Big[C[1^4] N_A^2 N_B^4+C[2,1^2]2 N_A \left(N_A^2+2\right) N_B^3\nonumber\\ &&\quad\qquad+C[2^2]N_A^2 \left(N_A^2+2\right) N_B^2+C[3,1]8 N_A^2 N_B^2\nonumber\\ &&\quad\qquad+C[4]2 N_A \left(2 N_A^2+1\right) N_B\Big]\nonumber\\
&&+ \Tr\Lambda_X^2 \Big[C[1^4]2 N_A \left(N_A^2+2\right) N_B^3\nonumber\\ &&\qquad\quad+C[2,1^2]2 N_A^2 N_B^2 \left(N_A^2+3 N_B^2+14\right)\nonumber\\ &&\qquad\quad+C[2^2]2 N_A N_B \left(N_A^2 \left(N_B^2+4\right)+2 \left(N_B^2+1\right)\right)\nonumber\\ &&\qquad\quad+C[3,1]8 N_A N_B \left(N_A^2 \left(N_B^2+2\right)+2 N_B^2+1\right)\nonumber\\ &&\qquad\quad+C[4]4 N_A^2 \left(N_A^2+8\right) N_B^2\Big]\nonumber\\ 
&&+ (\Tr\Lambda_X^2)^2
\Big[C[1^4]N_A^2 \left(N_A^2+2\right) N_B^2\nonumber\\ &&\qquad\quad+C[2,1^2]2 N_A N_B \left(N_A^2 \left(N_B^2+4\right)+2 \left(N_B^2+1\right)\right)\nonumber\\ &&\qquad\quad+C[2^2] N_A^2 N_B^2 \left(2 N_A^2+3 N_B^2+4\right)\nonumber\\ &&\qquad\quad+C[3,1]24 N_A^2 N_B^2\nonumber\\ &&\qquad\quad+C[4]2 N_A N_B \left(2 N_A^2 \left(N_B^2+1\right)+4 N_B^2+1\right)\Big]\nonumber\\
&&+ \Tr\Lambda_X^3  \Big[C[1^4]8 N_A^2 N_B^2\nonumber\\ &&\qquad\quad+C[2,1^2]8 N_A N_B \left(N_A^2 \left(N_B^2+2\right)+2 N_B^2+1\right)\nonumber\\ &&\qquad\quad+C[2^2]24 N_A^2 N_B^2\nonumber\\ &&\qquad\quad+C[3,1]8 N_A^2 N_B^2 \left(N_A^2+N_B^2+6\right)\nonumber\\ &&\qquad\quad+C[4]8 N_A N_B \left(N_A^2 \left(N_B^2+2\right)+2 N_B^2+1\right)\Big]\nonumber\\ 
%\left(\sum_{i=1}^{N}\lambda_i^3\right) 
&&+  \Tr\Lambda_X^4 \Big[C[1^4]2 N_A \left(2 N_A^2+1\right) N_B+C[2,1^2]4 N_A^2 \left(N_A^2+8\right) N_B^2\nonumber\\ &&\qquad\quad+C[2^2]2 N_A N_B \left(2 N_A^2 \left(N_B^2+1\right)+4 N_B^2+1\right)\nonumber\\ &&\qquad\quad+C[3,1]8 N_A N_B \left(N_A^2 \left(N_B^2+2\right)+2 N_B^2+1\right)\nonumber\\ &&\qquad\quad+C[4]2 N_A^2 N_B^2 \left(N_A^2+3 N_B^2+14\right)\Big]%\left( \sum_{i=1}^{N}\lambda_i^4\right) 
\Bigg\},\qquad 
 \end{eqnarray}
 from which it follows  
\begin{eqnarray}\label{eq:secMomNANB}
\fl {\cal M}_2^{A}(\Lambda_X)= c_{N_A,N_B}&&\Bigg[(N_B^2-1)(N_A^4N_B^2 (N_B^2-1) - 2 N_A^2(6N_B^2-7)+22)
\nonumber\\
\fl &&+ \Tr\Lambda_X^2 \; (2 N_A N_B (N_A^2-1)(N_B^2-1)(N_A^2N_B^2-14))\nonumber\\ 
\fl&&+ (\Tr\Lambda_X^2)^2
 \;(N_A^2-1)(N_B^4N_A^4+N_B^4N_A^2 - 14 N_A^2 N_B^2 + 6 N_B^2+30)\nonumber\\ 
 \fl &&+  \Tr\Lambda_X^3
 %\left(\sum_{i=1}^{N}\lambda_i^3\right) 
 \; 40 (N_A^2-1)(N_B^2-1)\nonumber\\ 
 \fl &&+ \Tr\Lambda_X^4
 %\left( \sum_{i=1}^{N}\lambda_i^4\right)
  \; (-10 N_A N_B)(N_A^2-1)(N_B^2-1)\Bigg], \qquad 
 \end{eqnarray}
 where
 \begin{eqnarray}\label{eq:COEFFsecMomNANB}
c_{N_A,N_B}=\frac{1}{N_A^2 N_B^2 (N_A^2 N_B^2-7)^2-36}.
 \end{eqnarray}
This expression generalizes the already known result for the pure case, when $\Tr \Lambda_X^k=1$ for all $k$ \cite{Zyczkowski3,Scott}.
 %which for the sake of simplicity is reported for the special case of $N_B=N_A=\sqrt{N}$.
In particular if $N_B=N_A=\sqrt{N}$  we get
 \begin{eqnarray}
{\cal M}_2^{A}(\Lambda_X)=&&\Bigg\{ \qquad\quad \Big[C[1^4] N^3+C[2,1^2] 2 N^2 (N+2)\nonumber\\ &&\quad\qquad+C[2^2]N^2 (N+2)+C[3,1]8N^2+C[4]2 N (2 N+1)\Big]\nonumber\\
&&+ \Tr\Lambda_X^2 \Big[C[1^4]2 N^2 (N+2)+C[2,1^2]4 N^2 (2 N+7)\nonumber\\ &&\qquad\quad+C[2^2]2 N (N (N+6)+2)\nonumber\\ &&\qquad\quad+C[3,1]8 N (N (N+4)+1)+C[4]4 N^2 (N+8)\Big]\nonumber\\ 
&&+ (\Tr\Lambda_X^2)^2 \Big[C[1^4]N^2 (N+2)+C[2,1^2]2 N (N (N+6)+2)\nonumber\\ &&\qquad\quad+C[2^2]N^2 (5 N+4)+C[3,1]24N^2+C[4]2 N (2 N (N+3)+1)\Big]\nonumber\\ 
&&+ \Tr\Lambda_X^3 %\left(\sum_{i=1}^{N}\lambda_i^3\right) 
\Big[C[1^4]8N^2+C[2,1^2]8 N (N (N+4)+1)\nonumber\\ &&\qquad\quad+C[2^2]24N^2 +C[3,1]16 N^2 (N+3)+C[4]8 N (N (N+4)+1)\Big]\nonumber\\
&&+  \Tr\Lambda_X^4 \Big[C[1^4]2 N (2 N+1)+C[2,1^2]4 N^2 (N+8) \nonumber\\ &&\qquad\quad+C[2^2]2 N (2 N (N+3)+1)+C[3,1]8 N (N (N+4)+1)\nonumber\\ &&\qquad\quad+C[4]4 N^2 (2 N+7)\Big]%\left( \sum_{i=1}^{N}\lambda_i^4\right) 
\Bigg\},\qquad 
 \end{eqnarray}
 that is 
 \begin{eqnarray}
\label{VVFIN2}
{\cal M}_2^{A}(\Lambda_X)= c_{N}&&\Bigg[(N^5 -2 N^4-11N^3+26N^2+8N-22)\nonumber\\
&&+ \Tr\Lambda_X^2 \; (2 N^5-4N^4-26N^3+56N^2-28N)\nonumber\\ 
&&+ (\Tr\Lambda_X^2)^2 \;(N^5-15N^3+20N^2+24 N -30)\nonumber\\ 
&&+ \Tr\Lambda_X^3 %\left(\sum_{i=1}^{N}\lambda_i^3\right) 
\; 40 (N-1)^2\nonumber\\
&&+  \Tr\Lambda_X^4 %\left( \sum_{i=1}^{N}\lambda_i^4\right) 
\; (-10 N )(N-1)^2\Bigg],\qquad 
 \end{eqnarray}
 with
 \begin{eqnarray}
c_{N}=\frac{1}{N^2 (N^2-7)^2-36}.
 \end{eqnarray}\\
 In the thermodynamical limit, $N>>1$, we find
 \begin{equation}
 {\cal M}_2^{A}=\frac{1}{N}(1+x)^2+\Ord{\left(\frac{1}{N^2}\right)}.
 \end{equation}

From Eqs. (\ref{eq:firstMom}), (\ref{eq:secMomNANB})  and (\ref{eq:COEFFsecMomNANB}) we can now compute the exact expression for the second cumulant of the purity at $\beta=0$:
\begin{eqnarray}\label{eq:secCumInZero}
  \mathcal{K}^A_2(x,0)&=& {\cal M}^A_2(x,0) - ({\cal M}^A_1(x,0))^2  \nonumber\\ 
 &=& +\frac{2(N_A^2 -1) (N_B^2 -1)( N_A^2 N_B^2+11) }{(N_A^2N_B^2-1)^2(N_A^4N_B^4-13 N_A^2N_B^2+36)}
 \nonumber\\ 
&&+x\;\frac{2(N_A^2 -1) (N_B^2 -1) (-2N_A N_B)(N_A^2 N_B^2+11)  }{(N_A^2N_B^2-1)^2(N_A^4N_B^4-13 N_A^2N_B^2+36)}
\nonumber\\ 
&&+ x^2\;\frac{2(N_A^2 -1) (N_B^2 -1) (N_A^4 N_B^4 - 4 N_A^2 N_B^2+15) }{(N_A^2N_B^2-1)^2(N_A^4N_B^4-13 N_A^2N_B^2+36)}\nonumber\\ 
 &&+ \langle \Tr\Lambda_X^3 \rangle_x  %\left(\sum_{i=1}^{N}\lambda_i^3\right) 
 \; \frac{40 (N_A^2-1)(N_B^2-1)}{N_A^2 N_B^2 (N_A^2 N_B^2 -7)^2-36}\nonumber\\  
 && + \langle \Tr\Lambda_X^4 \rangle_x % \left( \sum_{i=1}^{N}\lambda_i^4\right) 
 \; \frac{(-10 N_A N_B)(N_A^2-1)(N_B^2-1)} {N_A^2 N_B^2 (N_A^2 N_B^2 -7)^2-36}.
\end{eqnarray}
See Ref. \cite{Lubkin,Giraud} for the case $x=1$, when all the traces are 1.
%Assuming $N_B=N_A=\sqrt{N}$ we get
%\begin{eqnarray}
%  \mathcal{K}^A_2(x,0)&=& \;-\frac{2( N^2+11) }{(N+1)^2(N^4-13 N^2+36)}\nonumber\\ \nonumber \\ &&+x^2\;\frac{4N( N^2+11)  }{(N+1)^2(N^4-13 N^2+36)}\nonumber\\ \nonumber \\ &&\qquad\quad- x^4\;\frac{2(N^4 - 4 N^2 +15) }{(N+1)^2(N^4-13 N^2+36)}\nonumber\\ \nonumber\\ &&-  \left(\sum_{i=1}^{N}\lambda_i^3\right) \; 40 (N-1)^2\nonumber\\ \nonumber\\&&- \left( \sum_{i=1}^{N}\lambda_i^4\right) \; (-10 N )(N-1)^2.
%  \end{eqnarray}

\begin{figure}
\begin{center}
\includegraphics[width=0.6\textwidth]{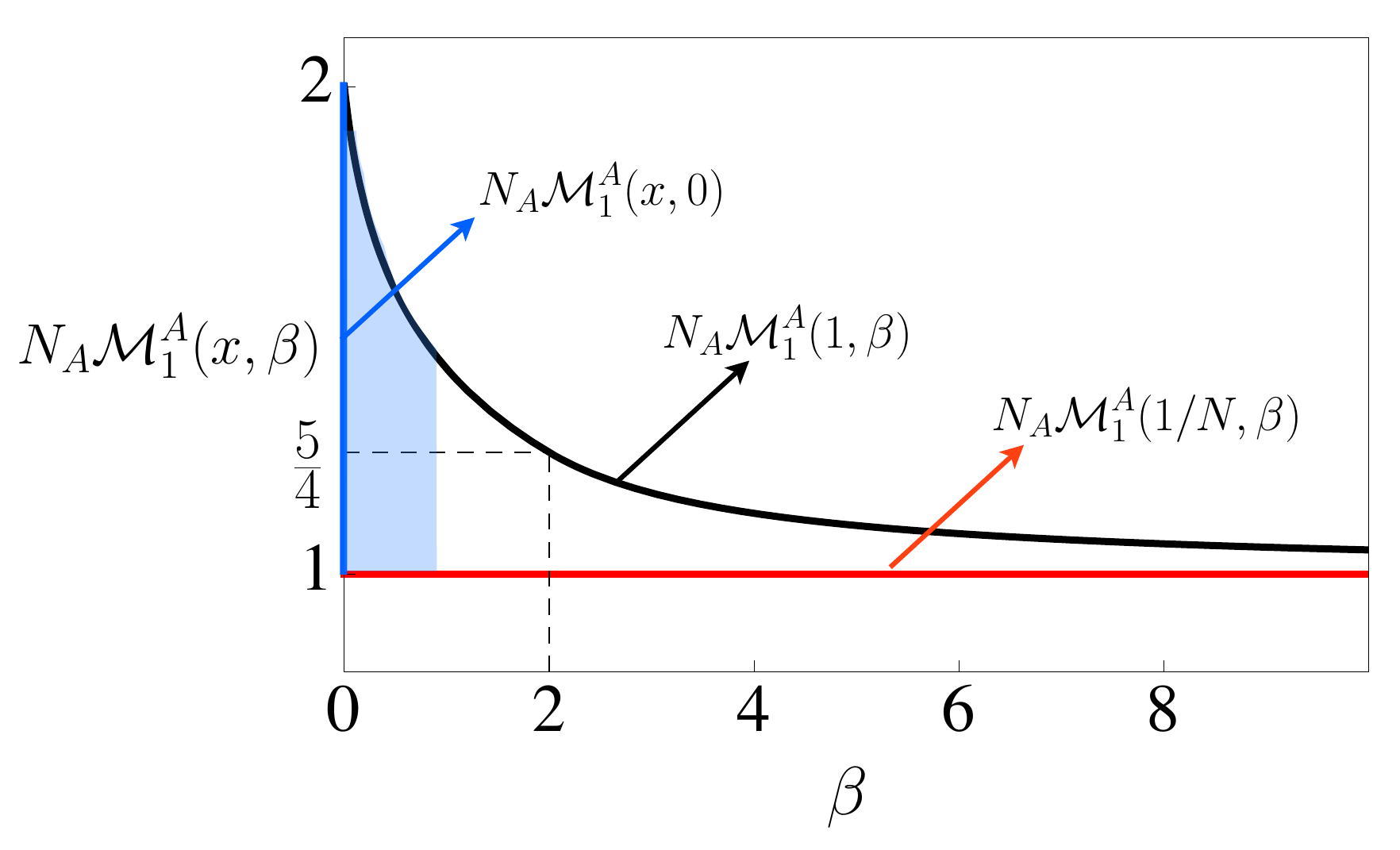}
\caption{
(Color online) First moment of the purity as a function of $\beta$. The horizontal (red) line refers to the set of totally mixed states, with $\mathcal{M}^A_1 (1/N, \beta)=1/N_A$; 
                                                                                                                   the (black) curve refers to pure states for $\beta > 0$; the vertical (blue) line corresponds to the first moment of the local purity at $\beta=0$, for arbitrary mixed states; finally,  
                                                                                                                   the shaded (light-blue) region refers to high temperatures.}
\label{fig:moment}
\end{center}
\end{figure}

\subsection{High temperature expansion of the first moment of the purity}
We can now compute the approximate expression for the first moment of the purity for small $\beta$, by plugging Eqs. (\ref{eq:primMomU}) and (\ref{eq:secCumInZero}) in Eq. (\ref {exp333}), with $n=2$: 
\begin{eqnarray}
{\cal M}_1^A(x,\beta)  &\sim&
 {\cal M}_1^A(x,0) + \beta \; [ ({\cal M}^A_1(x,0))^2 - {\cal M}^A_2(x,0)] \nonumber \\
&=& 
\frac{N_B (N_A^2 -1) }{N_A^2N_B^2-1} \;x +  \frac{N_A(N_B^2 -1) }{N_A^2N_B^2-1}\nonumber \\ 
&&+ \beta \; \Bigg[ \;-\frac{2(N_A^2 -1) (N_B^2 -1)( N_A^2 N_B^2+11) }{(N_A^2N_B^2-1)^2(N_A^4N_B^4-13 N_A^2N_B^2+36)}\nonumber\\ 
&&\qquad\quad-x\;\frac{2(N_A^2 -1) (N_B^2 -1) (-2N_A N_B)(N_A^2 N_B^2+11)  }{(N_A^2N_B^2-1)^2(N_A^4N_B^4-13 N_A^2N_B^2+36)}\nonumber\\ 
 &&\qquad\quad- x^2\;\frac{2(N_A^2 -1) (N_B^2 -1) (N_A^4 N_B^4 - 4 N_A^2 N_B^2+15) }{(N_A^2N_B^2-1)^2(N_A^4N_B^4-13 N_A^2N_B^2+36)}\nonumber\\ 
&&\qquad\quad- \langle \Tr\Lambda_X^3 \rangle_x %\left(\sum_{i=1}^{N}\lambda_i^3\right) 
\; \frac{40 (N_A^2-1)(N_B^2-1)} {N_A^2 N_B^2 (N_A^2 N_B^2 -7)^2-36}\nonumber\\ 
&&\qquad\quad- \langle \Tr\Lambda_X^4 \rangle_x %\left( \sum_{i=1}^{N}\lambda_i^4\right) 
\; \frac{(-10 N_A N_B)(N_A^2-1)(N_B^2-1)} {N_A^2 N_B^2 (N_A^2 N_B^2 -7)^2-36} \Bigg]. 
\end{eqnarray} 
For a balanced bipartition, $N_B=N_A=\sqrt{N}$, we get  
\begin{eqnarray}
{\cal M}_1^A(x,\beta)  &\sim&
 {\cal M}_1^A(x,0) + \beta \; [ ({\cal M}^A_1(x,0))^2 - {\cal M}^A_2(x,0)] \nonumber \\
&=& 
\frac{\sqrt{N}(1+x) }{1+N} \nonumber \\ 
&&+ \beta \; \Bigg[ \;-\frac{2( N^2+11) }{(N+1)^2(N^4-13 N^2+36)}\nonumber\\ 
&&\qquad\quad+x\;\frac{4N( N^2+11)  }{(N+1)^2(N^4-13 N^2+36)}\nonumber\\ 
 &&\qquad\quad- x^2\;\frac{2(N^4 - 4 N^2 +15) }{(N+1)^2(N^4-13 N^2+36)}\nonumber\\ 
 &&\qquad\quad-  \langle \Tr\Lambda_X^3 \rangle_x %\left(\sum_{i=1}^{N}\lambda_i^3\right) 
 \; \frac{40 (N-1)^2} {N^2  (N^2  -7)^2-36}\nonumber\\ 
 &&\qquad\quad- \langle \Tr\Lambda_X^4 \rangle_x %\left( \sum_{i=1}^{N}\lambda_i^4\right) 
 \; \frac{(-10 N )(N-1)^2} {N^2  (N^2  -7)^2-36} \Bigg], \label{exp3334}
\end{eqnarray} 
and in the thermodynamical limit 
\begin{equation}\label{eq:mom1betathermodyn}
{\cal M}_1^A(x,\beta)  \sim \frac{1+x}{\sqrt{N}}-\frac{2 \beta}{N^2} x^2 + \Ord \left(\frac{1}{N^{3/2}}\right) .
\end{equation}
One might wonder whether higher order cumulants follow a pattern similar to~(\ref{eq:mom1betathermodyn}).
Notice that (\ref{eq:mom1betathermodyn}) suggests a convergence radius for the high temperature expansion $\beta_c \sim N^{3/2} (1+x)/2 x^2$, which grows indefinitely when $x\to0$, in accordance with Eq.~(\ref{ffhh11}). See also Fig.~\ref{fig:moment}.

Equation~(\ref{eq:mom1betathermodyn}) 
at $x=1$ can be compared with the results of Ref.~\cite{matrix} where $\beta$ was replaced by the scaled quantity $\beta= \beta' N^{3/2}$. With this choice our expression yields 
\begin{eqnarray}
{\cal M}_1^A(1,\beta'N^{3/2})  &\sim&
 {\cal M}_1^A(1,0) + \beta' N^{3/2} \; [ ({\cal M}^A_1(1,0))^2 - {\cal M}^A_2(1,0)]
\nonumber \\ &\sim& (1  -  \beta') \;  \frac{2 }{\sqrt{N}} 
\;, \label{exp3335}
\end{eqnarray} 
 in perfect agreement with the behavior reported in Fig.~2 of Ref.~\cite{matrix}. 
Figure~\ref{fig:moment} yields an overview and summarizes our results.

\section{An alternative approach based on Twirling}\label{sec:tw}

In this section we will establish an interesting connection between our problem and the theory of quantum channels. In particular, we present an alternative approach for computing the moments ${\cal M}^{A}_n(x, 0)$ of Eq.~(\ref{defma}) which exploits  the properties of {\em twirling transformations}~\cite{tw1,tw2,tw3,BENNETT,BENNETT1}.

For explanatory purposes we start in Sec. \ref{sec:twirlingpure} by deriving a general expressions for 
the moments ${\cal M}^{A}_n(x=1, 0)$ associated with the case  in which $\rho$ is pure (i.e. the quantities ${\cal M}_n$ of Eq.~(\ref{moma}))
and verify that it yields the exact value given by Lubkin~\cite{Lubkin} for $n=1$.
The case of mixed states is then addressed in Sec. \ref{sec:Twirling_mixed} showing that for $n=1$ it reproduces the results of Sec.~\ref{sec:Zuber1}.

\subsection{Pure initial states}
\label{sec:twirlingpure}

Let us consider a fixed (normalized) pure state of the global system $X=AB$,
$|\psi_0\rangle_{X}$, and parametrize the pure states of $\cH_X$ as in (\ref{eq:unitaryrot}), i.e. $|\psi \rangle_{X} : = U_{X} |\psi_0\rangle_{X}$, with $U_X \in \mathcal{U}(\mathcal{H}_X)\simeq\mathcal{U}(N)$
distributed according to the Haar measure $\mathrm{d}\mu_H(U)$.
%Consider now its reduce density matrix of $A$ associated with Eq.~(\ref{eq:unitaryrot}), i.e.
%\begin{eqnarray}\label{Vdue}
%\rho_A&=& 
%\Tr_B [ |\psi\rangle_{AB}\langle \psi|] %=  \Tr_B [U_{AB} |\psi\rangle_{AB}\langle \psi| U^\dag_{AB}]\\
%= \sum_{\ell=1}^{N_B}    {_B\langle} \ell |U_{AB} |\psi\rangle_{AB} \langle \psi| U_{AB}^\dag | \ell\rangle_B\;,
%\end{eqnarray}
%with $\{ |\ell\rangle_B: \ell \in\{ 1,\cdots, N_B\}\}$ being an orthonormal basis of $B$.
Its local purity~(\ref{eq:purityN_A})  can be expressed as  
\begin{eqnarray}
\fl \quad 
\pi(\psi) 
=\sum_{1 \leq \ell, \ell' \leq N_B}
  \Tr\left(  {_B\langle} \ell |U_{AB} |\psi_0\rangle_{AB} \langle \psi_0| U_{AB}^\dag | \ell\rangle_B\;
   {_B\langle} \ell' |U_{AB} |\psi_0\rangle_{AB} \langle \psi_0| U_{AB}^\dag | \ell'\rangle_B \right)\label{Vtre}
  \nonumber \\
 \fl\quad
  = \sum_{1 \leq \ell, \ell' \leq N_B}
 {_{AB} \langle} \psi_0| \left( U_{AB}^\dag | \ell\rangle{_B\langle} \ell' |U_{AB} \right) |\psi_0\rangle_{AB}\;
 {_{AB}\langle} \psi_0| \left( U_{AB}^\dag | \ell'\rangle_B \langle \ell |U_{AB} \right) |\psi_0\rangle_{AB}\;,
   \end{eqnarray}
   where  $\{ |\ell\rangle_B\}$ is an orthonormal basis of $\cH_B$, and 
  % where in the last term we used 
  the cyclicity of the trace was used. 
  %to write the resulting expression   as a product of two expectation values on $|\psi\rangle_{AB}$. 
  We can recast this expression into a more compact form by doubling the Hilbert space, i.e. 
   adding   two auxiliary  copies $A'$ and $B'$  of $A$ and $B$, respectively. We get 
   \begin{eqnarray}\nonumber
\pi(\psi) &=&  \Tr \Big[ \left( U_{AB} \otimes U_{A'B'}\right)
 \Big( | \psi_0 \rangle_{AB}\langle \psi_0|  \otimes  |\psi_0 \rangle_{A'B'}\langle \psi_0|  \Big)
 \left( U_{AB}^\dag \otimes U_{A'B'}^\dag\right)
 \nonumber\\
 &&\qquad \quad \times \left(S_{B|B'} \otimes I_{AA'}\right) \Big] \label{Vpurity}\;,
    \end{eqnarray}
      where  the trace is  over all degree of freedom (i.e. $AA'BB'$),   $I_{AA'}$ is the identity operator on $AA'$, and 
\begin{eqnarray}
 S_{B|B'} : = \sum_{1 \leq \ell, \ell' \leq N_B}
 | \ell\rangle{_B\langle} \ell' | \otimes  | \ell'\rangle{_{B'}\langle} \ell| \;, \label{Vswap}
\end{eqnarray}
is the SWAP operator on $BB'$ -- this is the unitary, self-adjont transformation which, for all operators $\Theta_B$ and $\Upsilon_{B'}$,
gives 
\begin{equation}
S_{B|B'} (\Theta_B\otimes \Upsilon_{B'}) S_{B|B'} =  \Upsilon_{B} \otimes \Theta_{B'}.
\end{equation}
Remembering that first moment ${\cal M}_1^A(x=1,0) = {\cal M}_1$ of Eq.~(\ref{moma}) is obtained by averaging over all possible $U_X$, we can then write 
  %of $\Tr [ \rho_A^2(U)]$ with respect to $U$ can be written as
    \begin{eqnarray}
{\cal M}_1&=& %\left\langle \Tr [ \rho_A^2(U)] \right\rangle=
 \Tr \Big( {\cal T}^{(2)}( | \psi_0 \rangle_{X}\langle \psi_0|  \otimes  |\psi_0 \rangle_{X'}\langle \psi_0|  ) \;
 \left(S_{B|B'} \otimes I_{AA'}\right) \Big)\;, \label{Vmedia}
    \end{eqnarray}
    where $X=AB$, $X'=A'B'$,  and where  ${\cal T}^{(2)}$ is the Completely Positive, Trace Preserving (CPTP) {\em twirling} channel~\cite{tw1,tw2,tw3,boundent2}
    which transforms the operators $\Theta_{XX'}$ of $XX'$ into
  \begin{eqnarray}\label{VVTW}
{\cal T}^{(2)} (\Theta_{XX'} ) =  \int d \mu_H(U) \; \left( U_{X} \otimes U_{X'}\right)\; \Theta_{XX'} \;
( U_{X}^\dag \otimes U_{X'}^\dag).
    \end{eqnarray}
%    with $d\mu_H(U)$ being the Haar measure. 
This map plays  an important role in quantum information theory where it was
first introduced as a tool for
characterizing the distillability of bipartite entanglement~\cite{BENNETT1,BENNETT}.
It  has several properties
which  allows us to simplify the calculation.  For instance it is known
that ${\cal T}^{(2)}$  maps all the states of the system into (generalized) Werner states~\cite{tw1,tw2}. Furthermore it  is self-adjont -- i.e.
its description in Heisenberg picture coincides with ${\cal T}^{(2)}$.
In particular this last property can be used to rewrite~(\ref{Vmedia}) as
    \begin{eqnarray}\label{Vmedia11}
    {\cal M}_1
%\left\langle \Tr [ \rho_A^2(U)] \right\rangle
%&=&  \Tr \Big[( | \psi \rangle_{X}\langle \psi|  \otimes  |\psi \rangle_{X'}\langle \psi|  ) \; {\cal T}\left(S_{B|B'} \otimes I_{AA'}\right) ]  \\\nonumber
  &=& \Big( {_{X}\langle}  \psi_0|  \otimes  {_{X'} \langle}  \psi_0| \Big)\;
  {\cal T}^{(2)}\left(S_{B|B'} \otimes I_{AA'}\right) \;   \Big(| \psi_0 \rangle_{X}\otimes |\psi_0\rangle_{X'} \Big)
 \;.  \end{eqnarray}
 %(this is a simple consequence of the cyclicity of the trace and of the fact that the Haar measure is invariant under
 %h.c., i.e.  $d\mu(U) = d\mu(U^\dag)$).
 Explicit expressions for the action of ${\cal T}^{(2)}$ can be obtained by exploiting the symmetry of $d\mu_H(U)$. In particular it is possible to show that  ${\cal T}^{(2)} (\Theta_{XX'} )$ can be decomposed as
a linear combination of the projections on the symmetric and anti-symmetric subspaces of $XX'=ABA'B'$ (with respect to the bipartition $AB|A'B'$). Introducing then the SWAP operator
 which exchanges $X$ with $X'$ this can then be written as
 \begin{eqnarray}
\fl & & {\cal T}^{(2)} (\Theta_{XX'} ) =  \frac{N I_{XX'} - S_{X|X'}}{N(N^2-1)} \; \Tr \, \Theta_{XX'} 
+ \frac{N  S_{X|X'} - I_{XX'} }{N(N^2-1)}\;  \Tr ( S_{X|X'} \Theta_{XX'}) 
\label{Vsimply}
\\
\fl & &\qquad =  \frac{N \Tr \Theta_{XX'}  - \Tr( S_{X|X'} \Theta_{XX'})  }{N(N^2-1)} \; I_{XX'}
+ \frac{N  \Tr ( S_{X|X'} \Theta_{XX'})-\Tr \Theta_{XX'}
}{N(N^2-1)}\;  S_{X|X'}  \;,
\nonumber\\
\fl \label{VF}
\end{eqnarray}
where  $N=N_AN_B$ is the dimension of  ${\cal H}_{AB}$.  (Here
 $I_{XX'}$ is the identity operator while  $S_{X|X'}=S_{AB|A'B'}=S_{B|B'} \otimes S_{A|A'}$ is the SWAP which exchanges $AB$ with $A'B'$).

Thanks to this expression we can now easily compute the value of ${\cal M}_1$,
 either using Eq.~(\ref{Vmedia}) or Eq.~(\ref{Vmedia11}).
 Consider for instance the first approach.
  We have first  to compute the quantities $\Tr \, \Theta_{XX'} $ and $\Tr ( S_{X|X'} \Theta_{XX'})$
 with $\Theta_{XX'}$ being the operator $|\psi_0\rangle_{X}\langle \psi_0| \otimes |\psi_0\rangle_{X'}\langle \psi_0|$. This is
 \begin{eqnarray}
 \Tr (|\psi_0\rangle_{X}\langle \psi_0| \otimes |\psi_0\rangle_{X'}\langle \psi_0|) =1\;,
 \nonumber \\
 \Tr\left((S_{B|B'} \otimes S_{A|A'} )|\psi_0\rangle_{AB}\langle \psi_0| \otimes |\psi_0\rangle_{A'B'}\langle \psi_0|\right)= 1\;,
 \end{eqnarray}
% We have first of all to compute the quantities $\Tr_{12} [ \Theta_{12} ]$ and $\Tr_{12} [ S_{1|2} \Theta_{12}]$
% with $\Theta_{12}$ being the operator $S_{B|B'} \otimes I_{AA'}$. This is
% \begin{eqnarray}
% \Tr_{ABA'B'} [S_{B|B'} \otimes I_{AA'}] &=& \Tr_{BB'} [S_{B|B'}]  \Tr[ I_{AA'}] =
% N_B N_A^2\;, \nonumber \\ \nonumber
% \Tr_{ABA'B'} [(S_{B|B'} \otimes S_{A|A'} )(S_{B|B'} \otimes I_{AA'})] &=& \Tr_{BB'} [I_{BB'}]  \Tr[ S_{A|A'}] =
% N_B^2 N_A\;,
% \end{eqnarray}
 where in the second expression we used  the fact that
 $|\psi\rangle_{AB}\otimes |\psi\rangle_{A'B'}$ is invariant uder $S_{X|X'}$, i.e.
 $(S_{B|B'} \otimes S_{A|A'} )(|\psi\rangle_{AB}\otimes |\psi\rangle_{A'B'} )= |\psi\rangle_{AB}\otimes |\psi\rangle_{A'B'}$.
Replacing all this in Eq.~(\ref{Vsimply}) we get
   \begin{eqnarray}
{\cal T}^{(2)} (|\psi_0\rangle_{X}\langle \psi_0| \otimes |\psi_0\rangle_{X'}\langle \psi_0|) &=&  \frac{I_{XX'} + S_{X|X'}}{N(N+1)}  \;,
\end{eqnarray}
and thus
    \begin{eqnarray}
%\left\langle \Tr [ \rho_A^2(U)] \right\rangle
{\cal M}_1&=&\frac{1}{N(N+1)}
 \Tr \Big( ( I_{ABA'B'} + S_{B|B'}\otimes S_{A|A'})  \;
 \left(S_{B|B'} \otimes I_{AA'}\right) \Big) \nonumber \\ &=&
 \frac{1}{N(N+1)}  \Big(\Tr \Big( S_{B|B'} \otimes I_{AA'} \Big) +
\Tr \Big( I_{BB'} \otimes S_{A|A'} \Big)\Big)
 \; \label{Vmedia10}
    \end{eqnarray}
    (here we exploited the fact that $ S_{B|B'}^2=I_{BB'}$).
Now we can use $N=N_AN_B$ and the identities
\begin{eqnarray}
\Tr \left( S_{B|B'} \otimes I_{AA'} \right)  = N_B N_A^2\;, \quad
\Tr \left(  I_{BB'} \otimes S_{A|A'} \right)  = N_A N_B^2\;,
\end{eqnarray}
to get    \begin{eqnarray}
%\left\langle \Tr [ \rho_A^2(U)] \right\rangle =
{\cal M}_1=
 \frac{N_A + N_B}{N_AN_B+1}
 \;,\label{Vmedia20}
 \end{eqnarray}
 which coincides with the correct value~\cite{Lubkin,Giraud}.

 %\paragraph{Higher moments:--}
 %{\em NOTE: in this section we use a more friendly notation emitting the labels of the spaces.}

We mention that the same techniques can also be applied to higher moments ${\cal M}_n$. 
%Indeed for  $k$ integer  the quantities~(\ref{moma}) 
%     \begin{eqnarray}
%{\cal M}_k:= \left\langle \left( \Tr [ \rho_A^2(U)] \right)^k\right\rangle&=&  \int d \mu(U) \; \Big( \Tr [ \rho_A^2(U)] \Big)^k\;.
%%\nonumber \\
%\label{Mk}\end{eqnarray}
The extension  of Eq.~(\ref{Vmedia}) for $n\geqslant 2$ is obtained by
introducing $2n$ copies of $AB$ organized  in the  $n$ pairs, i.e.  $A_1 B_1 A_1'B_1'$, $A_2 B_2 A_2'B_2'$,
$\cdots$, $A_nB_n A_n' B_n'$.
We then introduce the following {\em generalized} twirling transformation acting on
$XX'=A_1 B_1 A_1'B_1'$, $A_2 B_2 A_2'B_2'$,
$\cdots$, $A_n B_n A_n' B_n'$, i.e.
  \begin{eqnarray}\label{Vgentw}
\fl\qquad {\cal T}^{(2n)} (\Theta_{XX'}) =  \int d \mu_H(U) \; ( \underbrace{U\otimes U \otimes \cdots \otimes U}_{2n}) \; \Theta_{XX'} \;
( \underbrace{U^\dag\otimes U^\dag \otimes \cdots \otimes U^\dag}_{2n}),
    \end{eqnarray}
  with $\Theta_{XX'}$ being a generic operator on ${\cal H}_{XX'}:= {\cal H}_{AB}^{\otimes 2 n}$.  
  This channel is  a proper generalization of the map ${\cal T}^{(2)}$ whose properties can be established along the lines of Ref.~\cite{Zuber}. 
  With this choice Eq.~(\ref{moma})  then can be expressed as
\begin{eqnarray}
{\cal M}_n^A(x=1,0)= {\cal M}_n&=& \Tr\Big( {\cal T}^{(2n)} \Big( |\Psi^{\otimes 2}\rangle\langle \Psi^{\otimes 2}|^{\otimes n} \Big) (S_{\cal B}^{(2n)} \otimes I_{\cal A}^{(2n)}) \Big)
 \label{Vmediak}\;,
    \end{eqnarray}
   where $|\Psi^{\otimes 2} \rangle^{\otimes n}: = \otimes_{j=1}^n\big( |\psi\rangle_{A_jB_j} \otimes  |\psi\rangle_{A_j'B_j'}\big)$, $I_{\cal A}$ being the
   identity on the $2n$ copies of $A$, i.e. ${\cal A}= A_1A_1' \cdots A_n A_n'$, and with $S_{\cal B}^{(2n)}$ being the  SWAP operator which exchanges $B_1B_2 \cdots B_n$ with $B_1' B_2' \cdots B_n'$ pairwise, i.e.
   $S_{\cal B}^{(2n)} = \otimes_{j=1}^{n} S_{B_j|B_j'}$.

\subsection{Mixed initial states}\label{sec:Twirling_mixed}
Consider now the case with $x<1$.  Following the parameterization introduced in Secs.~\ref{sec:introductionstatisticalapproach} and \ref{sec:moments}, we split the average
over the set $\mathfrak{S}_{x}({\cal H}_X)$ of the density matrixes of global purity $x$, as an average over the unitary rotations of  acting on ${\cal H}_{X}$ followed
by an average over the  space of the eigenvalues $\rho$ -- see Eq.~(\ref{eq:partition function unitary rotated}). Specifically this is accomplished by writing 
%\begin{eqnarray}
$\rho(U)= U_{AB} \;\Lambda_{AB}  \;U_{AB}^\dag\;,$
%\end{eqnarray}
with $U_{AB}$ being a generic unitary transformation on ${\cal H}_{AB}$,  while $\Lambda_{AB}$ represents a given arbitrary choice of the system spectrum, see Eq. (\ref{eq:fixedSpectrum}).
%\begin{eqnarray}\label{VVpuri}
%\Lambda_{AB} = \sum_{i=1}^{N_A}\sum_{\ell=1}^{N_B}  \; \lambda_{i,\ell} \; |i\rangle_{A}\langle i | \otimes
 %|\ell \rangle_{B}\langle \ell |
%\;,
%\end{eqnarray}
%with $\{ |i\rangle_{A}\}$,  $i=1,\cdots, N_A$ and  $\{ |\ell\rangle_{B}\}$, $\ell=1,\cdots, N_B$  being orthonormal bases
 %of ${\cal H}_{A}$ and ${\cal H}_B$, respectively.  
% [Notice that the maximum number of eigenvalues
%of  a density matrix $\rho_{AB}$ is $N_AN_B$. Therefore they could be labelled with a single index of $N_AN_B$ entries. Here we have chosen to parametrize it with  two  indexes of smaller dimensions, i.e.    $i$ and $\ell$]. 
For convenience, let us rewrite the purification  (\ref{eq.ref_state})  of the density matrix $\rho$ as 
\begin{eqnarray}
|\Psi\rangle_{ABab} = \sum_{1\leq \alpha \leq N_A} \sum_{1 \leq \beta \leq N_B} \sqrt{\lambda_{\alpha\beta}}\; \; | \alpha\rangle_A\otimes |\beta\rangle_B \otimes
|\alpha\rangle_a \otimes | \beta\rangle_b\;,
\end{eqnarray}
where now the ancillary systems, isomorphic to $A$ and $B$, are labelled by $a$ and $b$, respectively.
%One can easily verify that $U_{AB} | \Psi\rangle_{ABab}$ is a purification of $\rho_{AB}(U)$.
The reduced density matrix $\rho_A(U)= \Tr_B\rho(U)$ can thus be written as
\begin{eqnarray}
\rho_A(U) &=&
 \Tr_{Bab}(U_{AB} |\Psi\rangle_{ABab} \langle \Psi| U_{AB}^\dag) \nonumber \\
 &=& \sum_{1\leq q \leq N_A N_B^2} {_{Bab} \langle} q | U_{AB} |  \Psi\rangle_{ABab} \langle \Psi| U_{AB}^\dag
 | q \rangle_{Bab}\;,
 \end{eqnarray}
with $\{ |q\rangle_{Bab}\}$ being an orthonormal basis of $Bab$.
Similarly the local A-purity of $\rho_A(U)$ becomes
\begin{eqnarray}
\pi_A(\rho)= \Tr \rho_A^2(U)&=& \sum_{1 \leq q, q' \leq N_A N_B^2}
  \Tr\Big(  {_{Bab}\langle} q |U_{AB} |\Psi\rangle_{ABab} \langle \Psi| U_{AB}^\dag | q\rangle_{Bab} \nonumber \\
  && \qquad \times
   {_{Bab}\langle} q' |U_{AB} |\Psi\rangle_{ABab} \langle \Psi| U_{AB}^\dag | q'\rangle_{Bab} \Big)\nonumber \\
  & = &\label{Vtre11}
   \sum_{1 \leq q, q' \leq N_A N_B^2}
{_{ABab}  \langle} \Psi| U_{AB}^\dag | q\rangle_{Bab} {\langle} q' |U_{AB} |\Psi\rangle_{ABab}
\nonumber \\
&& \qquad \times {_{ABab}  \langle} \Psi| U_{AB}^\dag | q' \rangle_{Bab} {\langle} q|U_{AB} |\Psi\rangle_{ABab}
 \end{eqnarray}
which, once more, can be casted as an expectation value on $|\Psi\rangle^{\otimes 2}$ by doubling the space (see Eq. (\ref{eq:locpuritydoublespace})). 
 Therefore by integrating over $U_{AB}$ we get,
\begin{eqnarray}
{\cal M}_1^A(\Lambda_X)= \Tr
\Big( {\cal T}^{(2)}\Big( |\Psi\rangle_{Xx}\langle \Psi| \otimes |\Psi\rangle_{X'x'}\langle \Psi|\Big)
\;\; \Big( S_{Bab|B'a'b'} \otimes I_{AA'} \Big)\Big)  \nonumber \\
= \Big( {_{Xx}\langle} \Psi| \otimes  {_{X'x'}\langle} \Psi| \Big)\label{VIMPO1}
 {\cal T}^{(2)}\Big( S_{Bab|B'a'b'} \otimes I_{AA'} \Big) \Big( |\Psi\rangle_{Xx} \otimes |\Psi\rangle_{X'x'}\Big),
 \end{eqnarray}
where ${\cal T}^{(2)}$ being the twirling transformation on $XX'$ of Eq.~(\ref{VVTW}) with
 $X=AB$, $X'=A'B'$, $x=ab$, and $x'=a'b'$ (here $A',a',B',b'$ are the auxiliary copies
 of $A,a,B$, and $b$ respectively).
 The above expression  is
 the average purity of the subsystem $A$ computed for states $\rho$ having the same spectra $\Lambda_X$.

 To compute the above quantity this time we use the last identity of Eq.~(\ref{VIMPO1}).
According to Eq.~(\ref{VF}) we have to compute $\Tr \Theta_{XX'} $ and $\Tr ( S_{X|X'} \Theta_{XX'} )$ with $\Theta_{XX'}$ being the operator $S_{Bab|B'a'b'} \otimes I_{AA'}=S_{B|B'}
\otimes S_{b|b'}\otimes S_{a|a'} \otimes I_{AA'}
$.
That is
 \begin{eqnarray}
\fl  \qquad & & \Tr_{ABA'B'} (S_{Bab|B'a'b'} \otimes I_{AA'}) = \Tr_{ABA'B'} (S_{B|B'} \otimes S_{b|b'}
 \otimes S_{a|a'} \otimes I_{AA'})\nonumber \\ 
\fl \qquad & &\qquad = 
 \Tr  (S_{B|B'}) \;
 \Tr (I_{AA'})   \; S_{b|b'} \otimes S_{a|a'}  =  N_A^2 N_B \; S_{b|b'} \otimes S_{a|a'} \;,
 \end{eqnarray}
 \begin{eqnarray} 
\fl\qquad  & & \Tr_{ABA'B'} \left( S_{AB|A'B'}(S_{Bab|B'a'b'} \otimes I_{AA'})\right) =
  \Tr_{ABA'B'} \left( I_{BB'} \otimes S_{A|A'} \otimes S_{b|b'}
 \otimes S_{a|a'} \right)
 \nonumber\\ 
 \fl \qquad & & \qquad = 
  \Tr( I_{BB'} )
 \; \Tr(S_{A|A'}) \; S_{b|b'} \otimes S_{a|a'}
 =  N_B^2 N_A \; S_{b|b'} \otimes S_{a|a'}  \;.
 \end{eqnarray}
 Thus from Eq.~(\ref{VF}) we get
 \begin{eqnarray}
\fl\qquad  {\cal T}^{(2)}\Big( S_{Bab|B'a'b'} \otimes I_{AA'} \Big)
&=&  \frac{N_B (N_A^2 -1) }{N_A^2N_B^2-1} \; I_{BB'} \otimes
I_{AA'} \otimes S_{b|b'} \otimes S_{a|a'}  \nonumber \\
\fl\qquad & &+  \frac{N_A(N_B^2 -1) }{N_A^2N_B^2-1}  \;  S_{B|B'} \otimes S_{A|A'} \otimes S_{b|b'} \otimes S_{a|a'}  \;,
 \label{VF11}
\end{eqnarray}
where we used $N=N_AN_B$. Replace now this into Eq.~(\ref{VIMPO1}) and employ the identities
\begin{eqnarray}
& \Big( {_{Xx}\langle} \Psi| \otimes  {_{X'x'}\langle} \Psi| \Big)
 (I_{XX'} \otimes  S_{x|x'} )
 \Big( |\Psi\rangle_{Xx} \otimes |\Psi\rangle_{X'x'}\Big) 
 %= \sum_{1\leq \alpha \leq N_A} \sum_{1\leq \beta \leq N_B}\lambda_{\alpha \beta}^2 
 = \Tr\Lambda_{X}^2\;,&
 \nonumber \\
& \Big( {_{Xx}\langle} \Psi| \otimes  {_{X'x'}\langle} \Psi| \Big)
 (S_{X|X'} \otimes  S_{x|x'} )
 \Big( |\Psi\rangle_{Xx} \otimes |\Psi\rangle_{X'x'}\Big) = 1\;.&
\end{eqnarray}
The final result is thus
\begin{eqnarray}
{\cal M}_1^A(\Lambda_X)\label{Vfin}
&=& \frac{N_B (N_A^2 -1) }{N_A^2N_B^2-1} \Tr\Lambda_{X}^2+  \frac{N_A(N_B^2 -1) }{N_A^2N_B^2-1} \;
 \end{eqnarray}
which is Eq.(\ref{eq:primMomU}) and depends upon the spectrum $\Lambda_X$ only through its purity. By
averaging upon on $\Lambda_X$ while keeping fix $x$ gives us the
same result~(\ref{eq:firstMom}).

\section{Conclusions} \label{sec:concl}

From the results obtained in the previous sections one can infer that the same phenomenon of concentration of measure that occurs for the eigenvalues of the reduced density matrices of pure states \cite{matrix,metastable} occurs in the present case as well. Indeed, we observe that for large $N$ the leading order of all $k$-moments equals the $k$-th power of the first moment (\ref{eq:VVFIN}), that, for a balanced bipartition $N_A=N_B=\sqrt{N}$, reads
\begin{equation}
{\cal M}^{A}_1(x, 0)=\frac{(1+x)}{\sqrt{N}}+\Ord\left(\frac{1}{N^{3/2}}\right).
\end{equation}
This observation spurs from the calculation of the second moment (\ref{VVFIN2}):
\begin{equation}
{\cal M}^{A}_2(x, 0) =\frac{(1+x)^2}{N}+\Ord\left(\frac{1}{N^{2}}\right).
\label{eq:secondoMomento}\end{equation}
Therefore,
\begin{equation}
{\cal M}^{A}_2(x, 0)={\cal M}^{A}_1(x, 0)^2+\Ord\left(\frac{1}{N^2}\right).
\end{equation}
By observing that the contributions of $\Ord(1/N)$ to the second moment  come from the coefficients proportional to $C\left[1^4\right]$ in (\ref{eq:kMomUnitIntl}), i.e.\  the identity permutation $\sigma=0$, and conjecturing that the leading contribution for all the $k$-th moments comes \emph{only} from $C\left[1^k\right] \sim 1/N^k$ one gets
\begin{equation}
{\cal M}^{A}_n(x, 0)= \left(\frac{(1+x)}{\sqrt{N}}\right)^k+\Ord\left(\frac{1}{N^{(k+2)/2}}\right).
\label{eq:stimamom_n}
\end{equation}
Another check of the validity of (\ref{eq:stimamom_n})  derives by the interpolation between  maximally mixed and pure global states. See Fig.~\ref{fig:moment}. The scaling with $N$ is preserved, therefore allowing to interchange the $x\to 1$ and $N\to\infty$ limits.

This uncovers the issue of computing the subdominant terms in the $k$-th moments expansion that become instead \emph{leading} order terms in the $k$-th \emph{cumulants}. This could be possible if we had an appropriate asymptotic expansion of the combinatorics coefficients $C$, which we do not know. We leave  this as a challenge for future work.

%%%%%%%%%%%%%%%%%%[
%\ack 

%%%%%%%%%%%%%%%%

\end{document}